\documentclass[a4paper,11pt]{article}

\usepackage{jinstpub}
\usepackage{graphicx}
\usepackage{xcolor}
\usepackage{caption}
\usepackage{subcaption}
\usepackage{amsmath}
\usepackage{amssymb}
\usepackage[numbers,sort&compress]{natbib}

\title{Study on the linearity of 20" dynode and MCP PMTs}

\author[a,b]{Diru Wu,}
\author[a,d]{Fengjiao Luo}
\author[a,c,e,1]{Zhimin Wang,\note{Corresponding author.}}
\author[a,c]{Min Li,}
\author[a,c,e]{Jilei Xu,}
\author[a,d,e]{Miao He,}
\author[a,c]{Changgen Yang,}
\author[a,c,e]{Yuekun Heng}

\affiliation[a]{Institute of High Energy Physics, Beijing 100049, China}
\affiliation[b]{China Center of Advanced Science and Technology, Beijing 100190, China}
\affiliation[c]{University of Chinese Academy of Sciences, Beijing 100049, China}
\affiliation[d]{University of South China, Hengyang 421001, Hunan, China}
\affiliation[e]{State Key Laboratory of Particle Detection and Electronics}

\emailAdd{wangzhm@ihep.ac.cn}

\abstract{The linearity of charge response is an important feature of photomultiplier tubes (PMT) for physics measurements, especially the newly developed 20" MCP-PMT. In this paper, in addition to the traditional method of double light sources, we applied another relative method of 20" PMT to 3" PMT to measure the linearity of the 20" dynode and MCP PMTs in pulse mode with a waveform digitizer. The measurements show a good linear response of 20" PMTs to 1,000 photoelectrons (p.e.). The correlations of the amplitude, rise-time, fall-time, FWHM, baseline recovery, overshoot, and late-pulse to the output charge of the 20" PMTs derived from the waveform analysis, where the MCP-PMT shows very different features compared to the dynode-PMT in particular.}


\arxivnumber{1234.56789} 

\date{December 2021}

\begin{document}

\maketitle
\flushbottom

\section{Introduction}
\label{1:intro}

Photomultiplier tubes (PMTs) are commonly used photon sensors sensitive to a single photon with many advantages, including low noise level, high dynamical range, good photon detection efficiency (PDE), attractive cost, and coverage ratio. PMTs are used in many particle physics experiments all over the world, especially in neutrino experiments, such as Double Chooz \cite{ChoozPMT, doubleChoozPMT, doubleChoozPMT2}, Super-Kamiokande \cite{Super-Kamiokande:1998uiq, PhysRevD.83.052010, SKPMTaccident}, Daya Bay \cite{dayabay, DayabayPMT}, LHAASO \cite{LHAASOPMT}, and JUNO \cite{JUNOdetector}. In the experiments, we need to measure the number of photons received by PMTs through their relationship between the number of photons and the measured charges. It is important to have a good understanding of the PMT linearity or non-linearity response for a precise charge measurement.

The linearity of PMT anode output can be measured in Direct Current mode (DC mode) or in pulse mode. The pulse mode within a certain frequency is more realistic compared to the signal features in the case of particle physics experiments, especially for the non-linearity effect studies. The photoelectron (p.e.) of PMT anode output per pulse, rather than the anode output current, is used to model the linearity response here.

There are already lots of studies and a good understanding of the linearity response of dynode PMTs \cite{PMTbook-photonis,HamManual}. The current linearity of PMT anode output is limited to 10\% of its working direct current of the high voltage divider (HV divider) from its gain response\cite{photonis-PMT-basic}. There are also measurements with the traditional methods of double light sources or different distances by some recent setup\cite{LHAASO-ZHANG2020162079,LHAASO-2019ICRC,LHAASO-KM2A-LV201534-2015}. The microchannel plate (MCP) PMT (MCP-PMT) incorporates an MCP in place of conventional discrete dynodes\cite{HamManual}. The response of the newly developed 20" MCP-PMT is also measured in \cite{optimization-PMT-signal-2018,JUNOPMTlinearity,MCP-charactor}. But it is still valuable to have a further study on methods and check more parameters about the PMT response linearity, related to pulse shape in particular.

Following the linearity measurement strategy discussed in \cite{JUNO-PMT2022}, the relative measurement method will be studied in detail, which will be compared with the results of the traditional double-light source method. At the same time, with the acquisition of waveform data, the pulse shape-related parameters, such as amplitude, rise-time, fall-time, full width at half maximum (FWHM), overshoot, and late-pulse, are also discussed. Besides, the newly developed 20" MCP-PMT is compared with the dynode-PMT.

Sec.\ref{1:setup} will provide the system setup and linearity measurement methods. Sec.\ref{1:results} will discuss the measured results, and a summary is collected in Sec.\ref{1:sum}.

\section{Experimental setup}
\label{1:setup}

\begin{figure}[!htpb]
	\centering
	\includegraphics[width=0.75\textwidth]{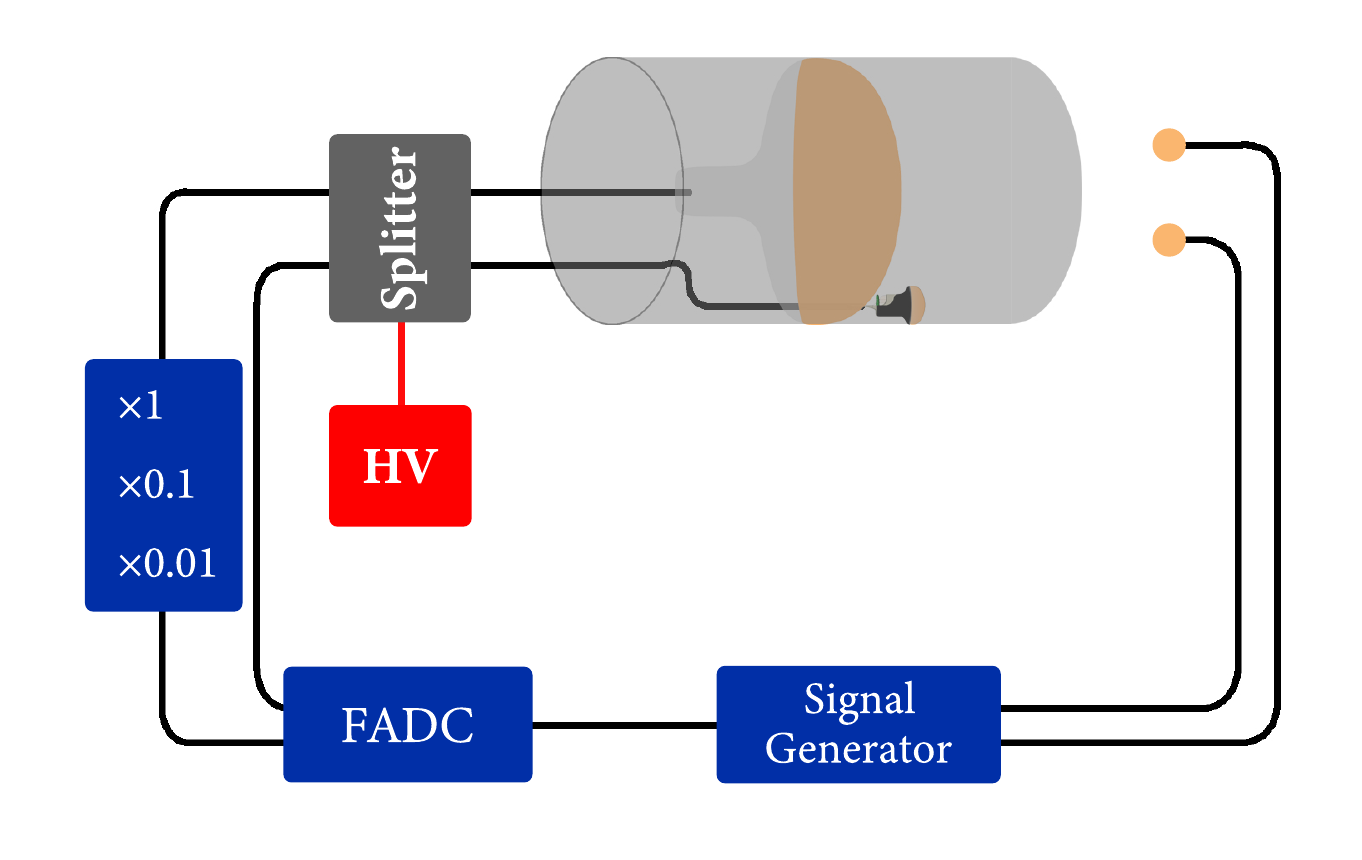}
	\caption{The layout of the linearity test system. The grey cylinder surrounding PMTs is an EMF shield to avoid the EMF effect on the 20" PMTs. The splitter is a circuit to de-couple the positive high voltage and signal of each PMT.}
	\label{fig:sys}
\end{figure}

A test system has been set up as shown in Fig.~\ref{fig:sys} where two LEDs, a 20" PMT (MCP-PMT or dynode-PMT) and a 3" PMT are used. The two LEDs, encased in a diffuser ball, were working in pulse mode and driven by a pulse generator which provides two synced pulses to flash the LEDs and another NIM (Nuclear Instrumentation Modules) pulse to trigger the digitizer. The digitizer is a 10-bit ADC in 0-1\,V range (CAEN DT5751, 1\,GSample/s). To reach a better signal-to-noise ratio and cover a larger dynamic range, an attenuator is used in the high-intensity cases with 0-2\,GHz bandwidth and 0.1 or 0.01 attenuation factor. A cylinder shielding surrounded the 20" PMT to avoid the Earth magnetic field (EMF) effect \cite{Zhang_2021-JUNO-EMF}.
The PMTs tested here, with positive high voltage, are 20" MCP-PMT GDB-6201 produced by NNVT (North Night Vision Technology Co.~Ltd), 20" dynode-PMT R12860 produced by HPK, and 3" dynode-PMT XP72B20 produced by HZC\cite{JUNO-PMT2022,JUNO3inchPMT}.

\begin{figure}[!htpb]
	\centering
	\includegraphics[width=0.75\textwidth]{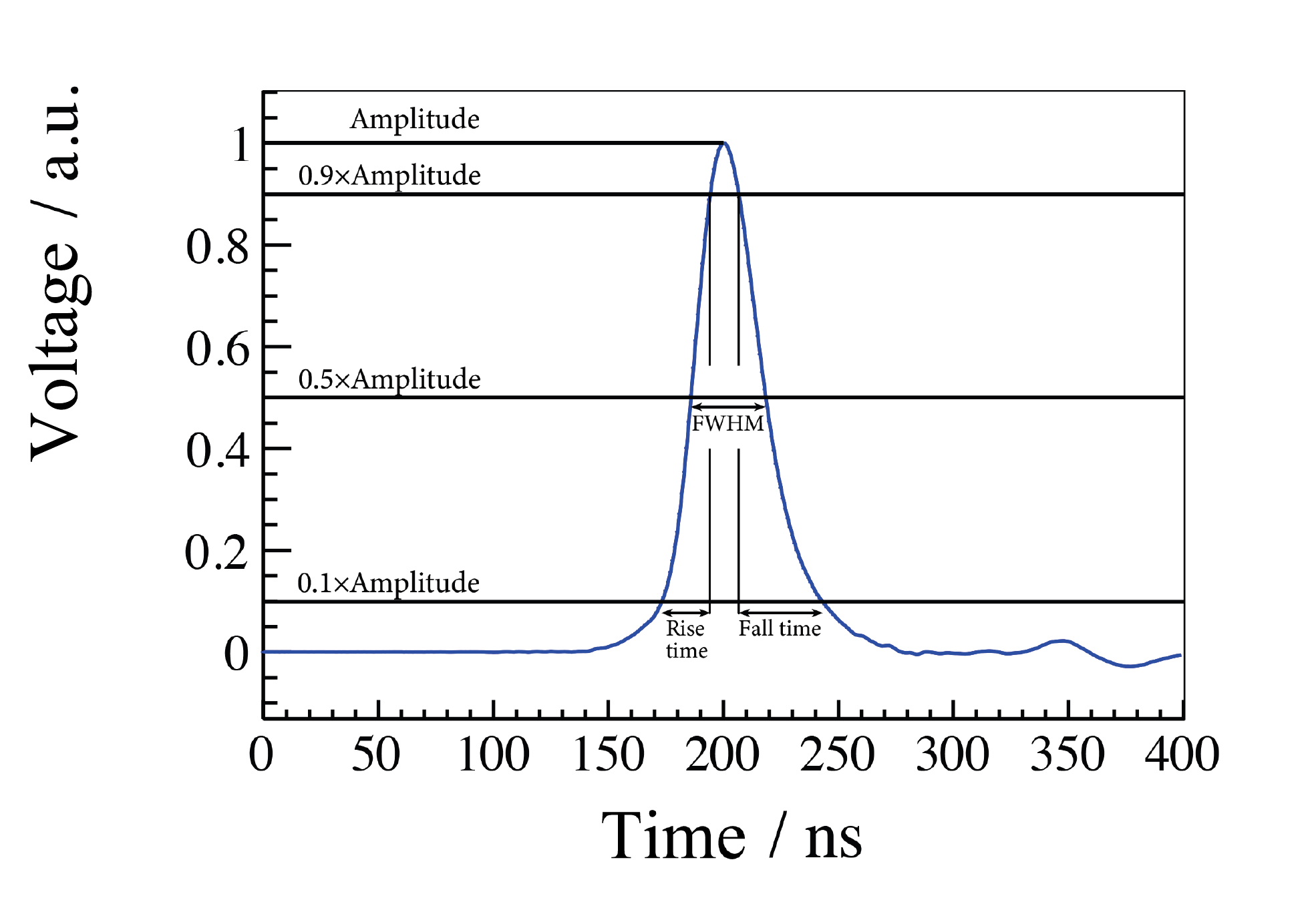}
	\caption{Definitions of PMT pulse-related parameters.}
	\label{fig:defs}
\end{figure}

The definitions of the pulse-related parameters are shown in Fig.~\ref{fig:defs}, where the pulse is inverted to positive. The amplitude in mV is the maximum value of the output pulse, and the rise-time and the fall-time are the time from $0.1\times$\,amplitude to $0.9\times$\,amplitude in the rising edge and the falling edge, respectively.
The charge of a PMT pulse is extracted by an integration of its waveform after considering the amplification factor and the gain of the PMT as in Eq.~\ref{eq:inte}, where the $Q$, $N$, $V_i$, $\Delta t$, 50~$\Omega$, $A$ and $G$ are the charge in photoelectron (p.e.), the number of the sampling points, the amplitude of the $i$-th sampling point, the sampling interval time, the impedance of the signal output, the amplification factor and the PMT gain. To cover the whole range of the linearity test with different pulse features, the charge integration was done in a fixed window of 600\,ns for 20" PMT ([-100,+500]\,ns relative to the peak of the primary pulse), the range was 200\,ns for 3" PMT ([-50,+150]\,ns relative to the peak of the primary pulse). But, the integration range was set to 100\,ns when the expected charge was less than 10\,p.e.~viewed by the 3" PMT to reduce the effect of baseline noise. As known, the results of charge calculation are affected by the overshoot and the late pulse, but the definition is still a good parameter for this study. The amplitude of the late pulse is derived in a range of [+80, +180]\,ns relative to the peak location of the primary pulse, and the amplitude of the overshoot is derived in a range of [+10, +200]\,ns relative to the peak location of the primary pulse. The baseline-recovery time is defined as the time from the peak of the primary pulse to the baseline back to zero after the primary pulse. The mean of each parameter in the measurement is selected as the checking value. During the tests, all the tested 20" PMTs are calibrated to $1\times10^7$ gain at first. 


\begin{equation}
	Q=\sum_{i}^{N}\frac{V_i \Delta t}{50 \Omega} \times \frac{1}{A} \times \frac{1}{G} 
	\label{eq:inte}
\end{equation} 

\begin{figure}[!htpb]
	\centering
	\includegraphics[width=0.75\textwidth]{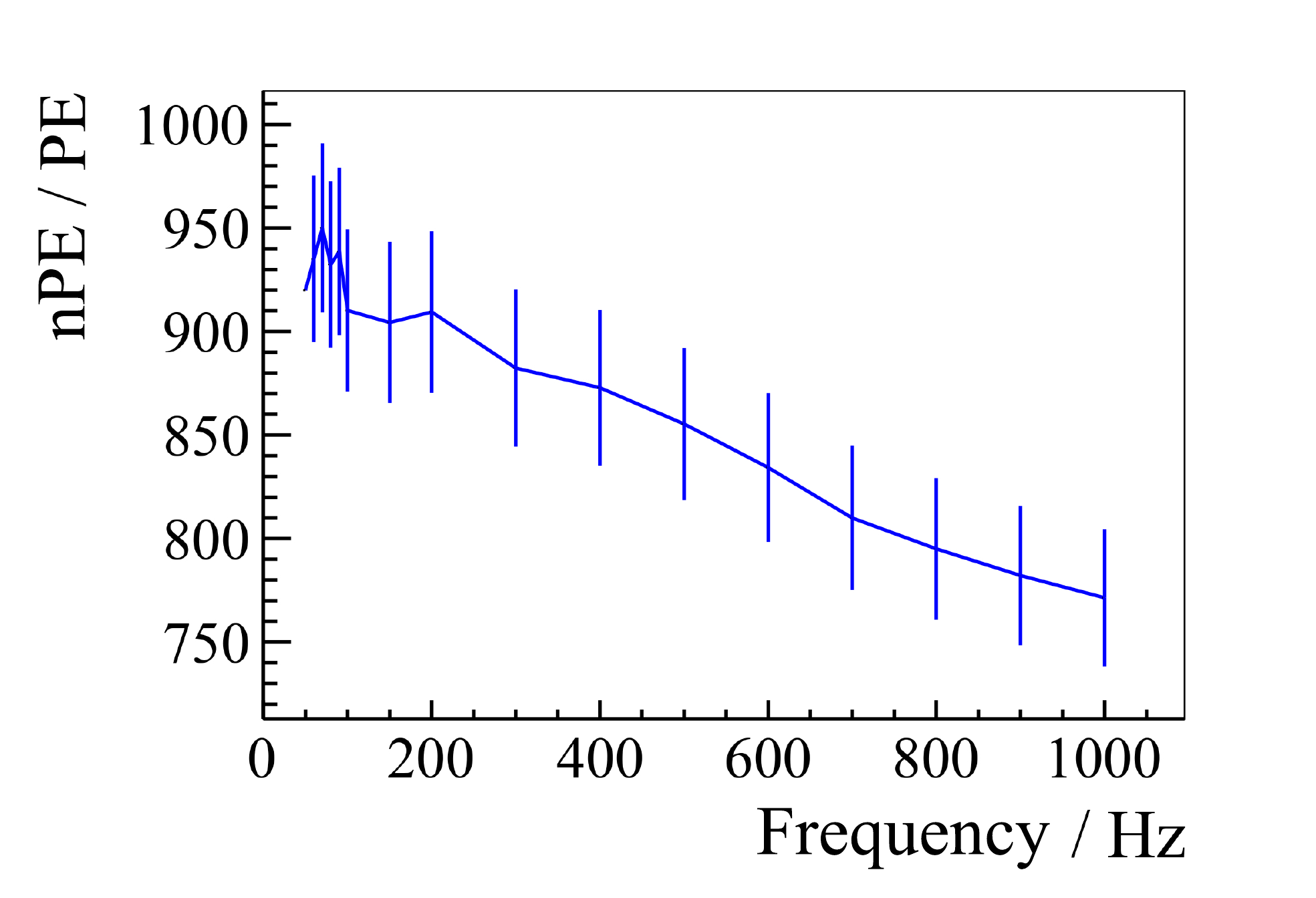}
	\caption{The charge output per pulse of the MCP-PMT to a constant intensity but the different frequencies of LED flashing. The error is only from the statistical uncertainty.}
	\label{fig:ResFre}
\end{figure}

As known, the linearity of the PMT response is related to the DC of the HV divider. The working current of PMT HV divider used in JUNO is around 5\,\textmu A for a gain $3\times10^6$ of 3" PMT \cite{JUNO3inchPMT,diru-2022arXiv220402612W}, around 90\,\textmu A for a gain of $1\times10^7$ of 20" dynode-PMT and around 180\,\textmu A for a gain of $1\times10^7$ of 20" MCP-PMT \cite{JUNOdetector,JUNOEmbedded,JUNOPMTsignalopt}, respectively. A 20" MCP-PMT (gain $1\times10^7$), as an example, is tested with a LED flashing in the constant intensity of a driver pulse width of 30\,ns but the different frequencies. The results are shown in Fig.~\ref{fig:ResFre}. It is obvious that the output charge per pulse of the PMT is decreasing anti-correlated to the LED frequency increasing as expected, which decreases from around 940\,p.e./pulse at 100\,Hz (1.5\,nC/pulse or 0.15\,\textmu A on average) to around 780\,p.e./pulse at 1\,kHz (1.2\,nC/pulse or 1.2\,\textmu A on average). But the equivalent current on average is still far from the limitation of about 10\% of the HV divider DC. Considering this effect and efficient data acquisition, a frequency of 100\,Hz with 30\,ns pulse width is chosen as the setting of the LEDs. During the tests, the amplitude of the driver pulse can be adjusted to get different light intensities.

\subsection{Double-LED method}
\label{2:2LEDs}

\begin{figure}[!htpb]
	\centering
	\includegraphics[width=0.4\textwidth]{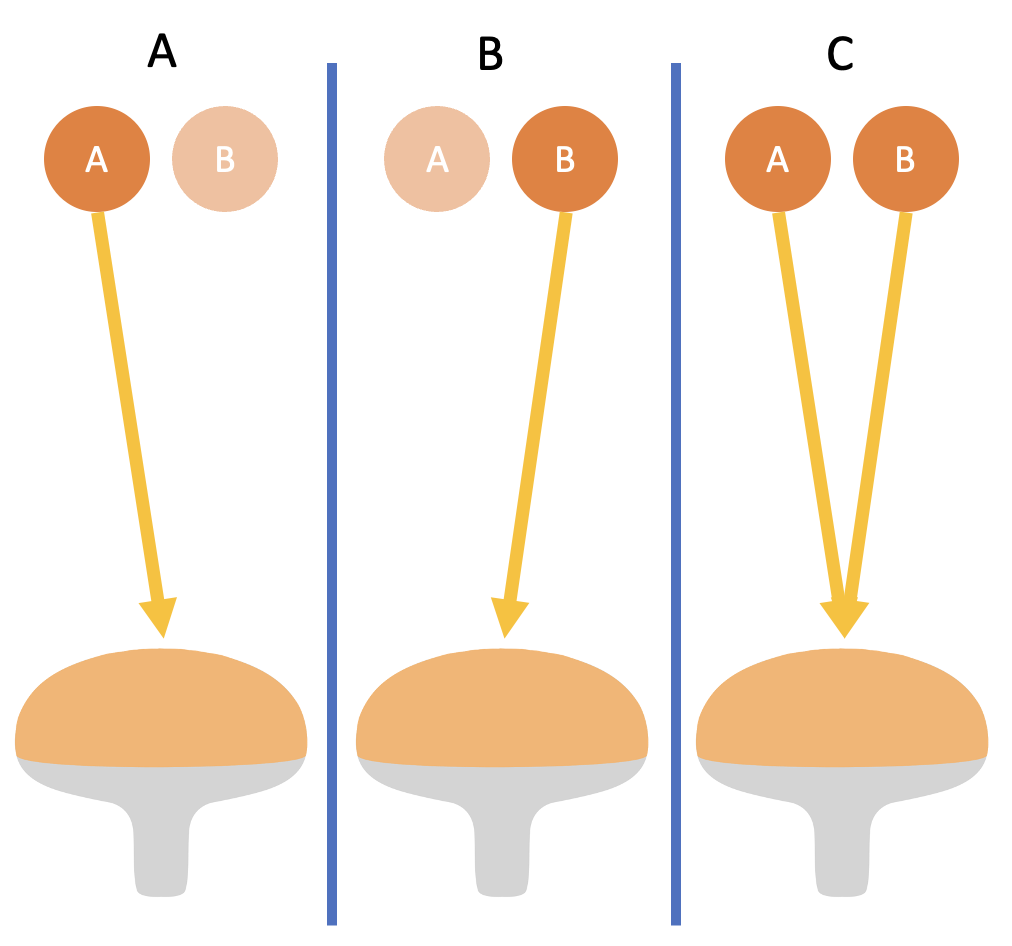}
	\caption{Schema of the double-LED method.}
	\label{fig:tradMethods}
\end{figure}

Traditionally, two light sources (LED used here) are commonly used to calibrate the linearity of PMT response, where the output charge of the PMT will be measured in a specified flashing sequence of the LEDs as shown in Fig.~\ref{fig:tradMethods}: $A$ flashing with LED "A", $B$ flashing with LED "B", and $C$ flashing with LED "A" and "B" at the same time (less than 2\,ns identified by the PMT rising time). The process will iterate to cover a required intensity range and derive the linear response. The nonlinearity of PMT of the measurement can be defined as $\frac{C-(A+B)}{A+B}$, and the nonlinearity of $A$ and $B$ also needs to be considered\cite{JUNOPMTlinearity}. Generally, this method needs a good configuration and lots of data-taking points to ensure precise measurement and is hard to realize.

\subsection{Relative method}
\label{2:relative}

\begin{figure}[!htpb]
	\centering
	\includegraphics[width=0.2\textwidth]{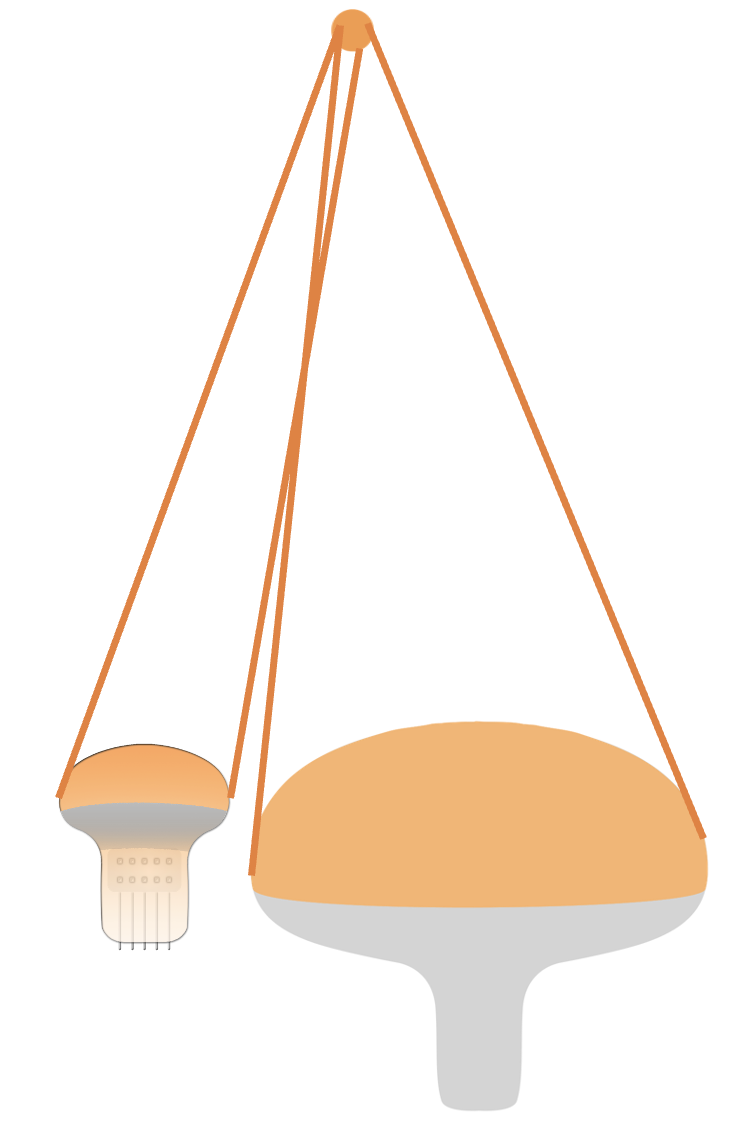}
	\caption{The relative method.}
	\label{fig:relaMethod}
\end{figure}

With a uniform light source and a similar distance, a 3" PMT can only collect around 1/40 photons viewed by a 20" PMT because of the solid angle (Fig.~\ref{fig:relaMethod}). In this case, it is possible to measure the linearity response of the 20" PMT by the 3" PMT, which still works in a linear range, as a relative method, which is useful for possible measurement as JUNO\cite{JUNOdetector}. This method can reduce the system complexity and the amount of data, and improve the operation efficiency.

\begin{figure}[!htpb]
	\centering
	\begin{subfigure}[c]{0.48\textwidth}
	\centering
	\includegraphics[width=\textwidth]{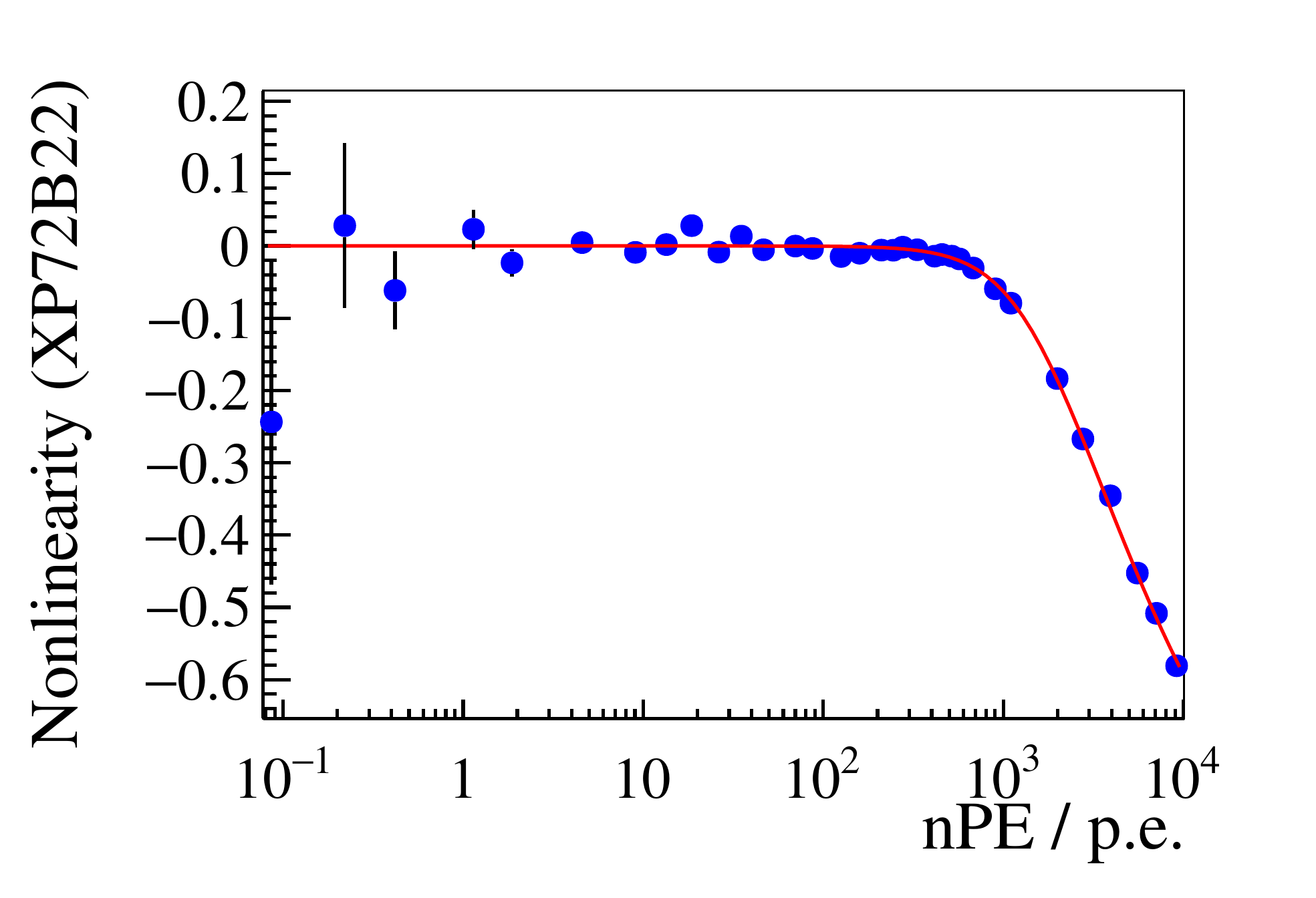}
	\caption{Nonlinearity of the 3" PMT.}
	\label{fig:linearity}
	\end{subfigure}	
	\begin{subfigure}[c]{0.48\textwidth}
	\centering
	\includegraphics[width=\textwidth]{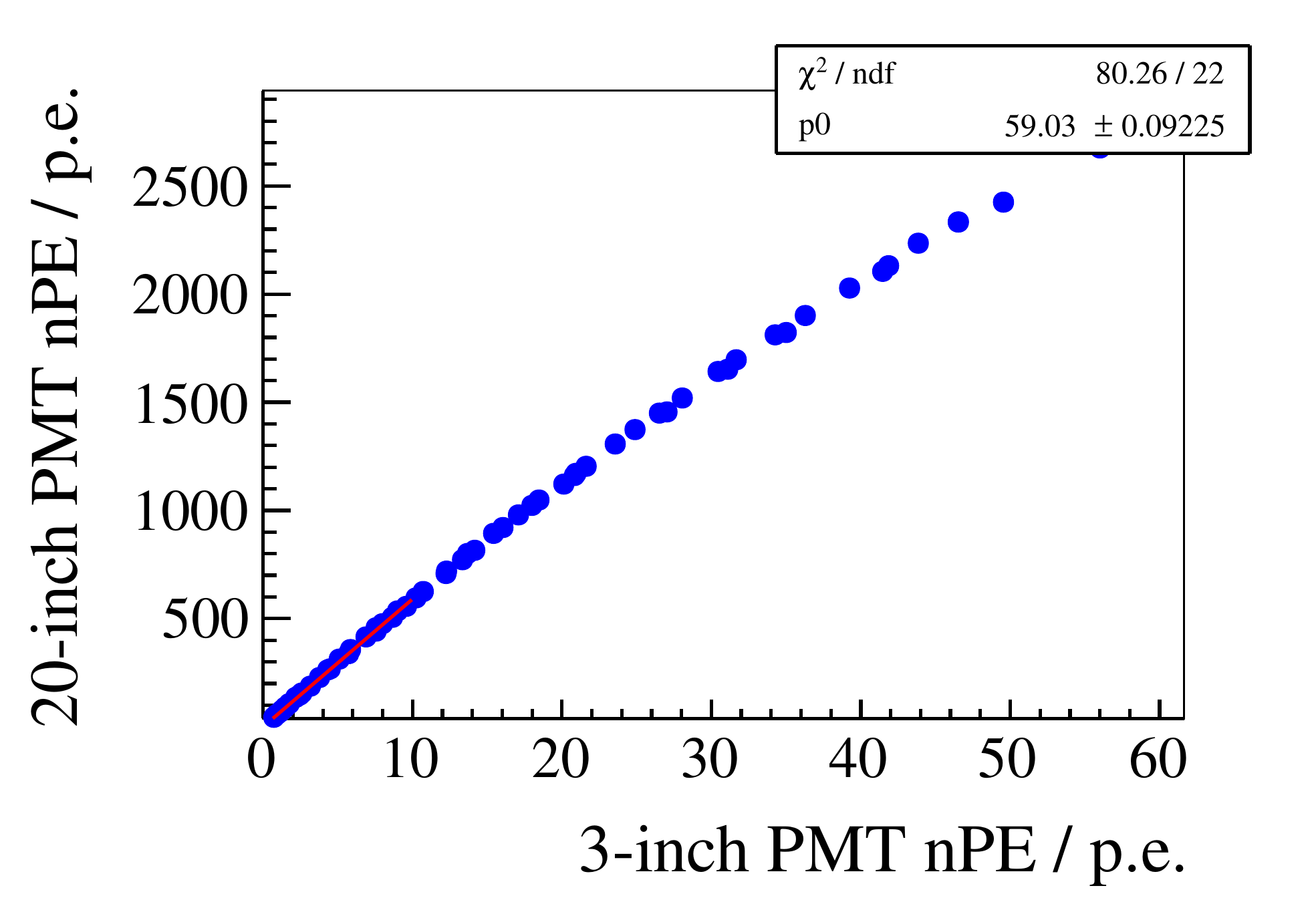}
	\caption{Intensity factor of the 3" PMT to 20" PMT.}
	\label{fig:3":factor}
	\end{subfigure}	
    \caption{(a) Linearity of the 3" PMT. (b) Intensity factor between the 20" and the 3" PMTs.}
	\label{fig:3inch:linearFitting}       
\end{figure}

The linearity response (nonlinearity factor) of the 3" PMT is first measured with the double-LED method as shown in Fig.\,\ref{fig:linearity}, where the points with small charge are affected a lot by the system noise where it is comparable with the signal intensity. 
The nonlinearity curves are further fitted by Eq.~\ref{eq:nonlinearity}\cite{PMTModel}, where $n$, $N_{pe}$ and $N_{pe}^{sat}$ are the ratio of non-linearity, the expected photoelectrons, and the position of nonlinearity $1-\frac{1}{\sqrt{2}}$, respectively. The range of the 3" PMT linearity response can reach around 750\,p.e.~as designed. 
A linear fitting is applied to the relative intensity curve measured by the 3" and 20" PMTs as shown in Fig.\,\ref{fig:3":factor}. A coefficient between the outputs of the 3" PMT and the 20" PMT is extracted (here is 59.03) to estimate the expected intensity of the 20" PMT, where the fitting range is limited to less than 20\,p.e.~viewed by the 3" PMT to avoid any nonlinearity effect of the 20" PMT.

\begin{equation}
	n=\sqrt{\frac{\sqrt{1+8\left( \frac{N_{pe}}{N_{pe}^{sat}}\right)^\alpha}-1}{4\left( \frac{N_{pe}}{N_{pe}^{sat}}\right)}}-1
	\label{eq:nonlinearity}
\end{equation}



\section{Linearity response}
\label{1:results}

\subsection{Charge}
\label{2:charge}

\begin{figure}[!htpb]
	\centering
	\begin{subfigure}[c]{0.45\textwidth}
	\centering
	\includegraphics[width=\textwidth]{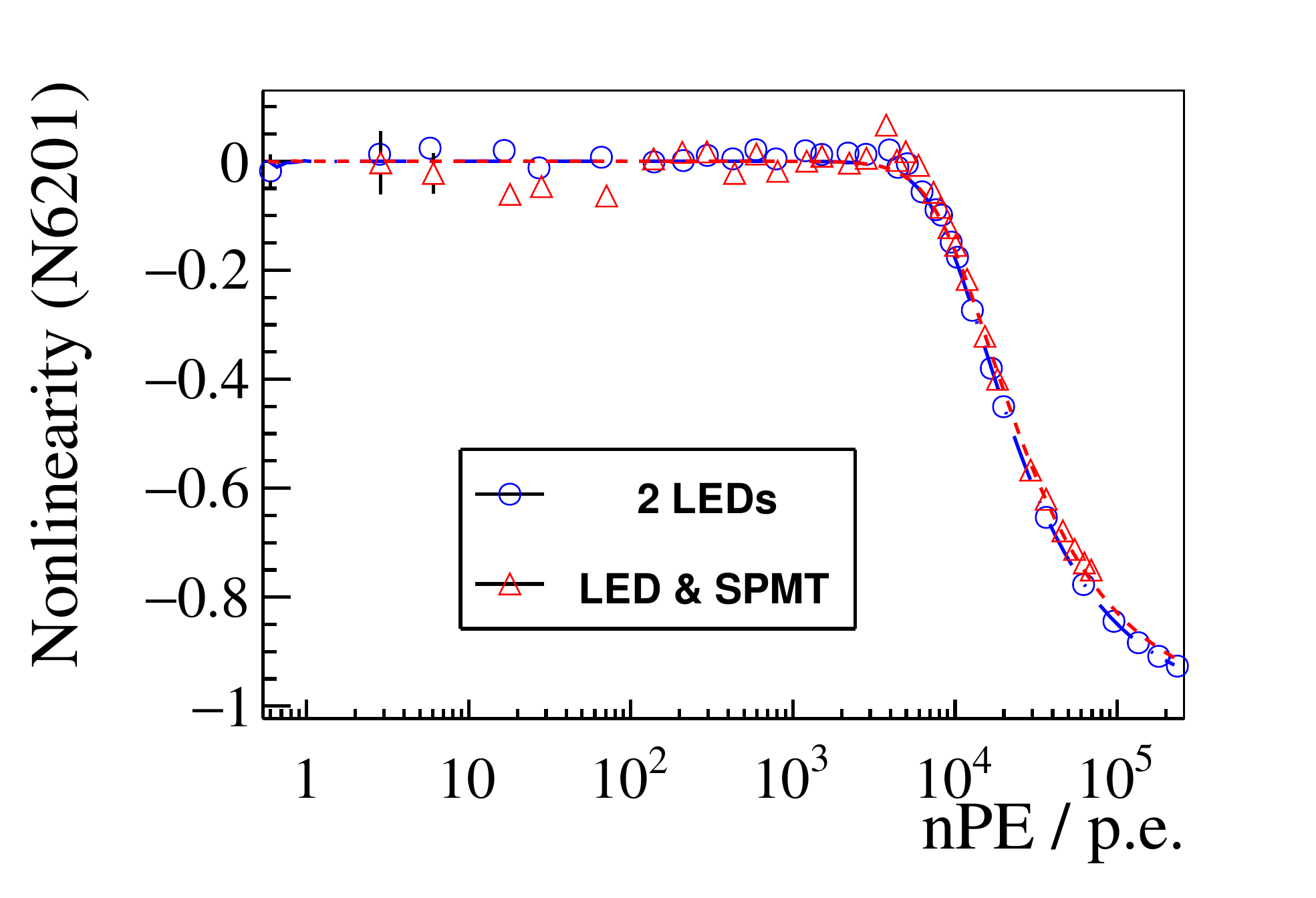}
	\caption{20" MCP-PMT}
	\label{fig:Charge:MCP}
	\end{subfigure}	
	\begin{subfigure}[c]{0.45\textwidth}
	\centering
	\includegraphics[width=\textwidth]{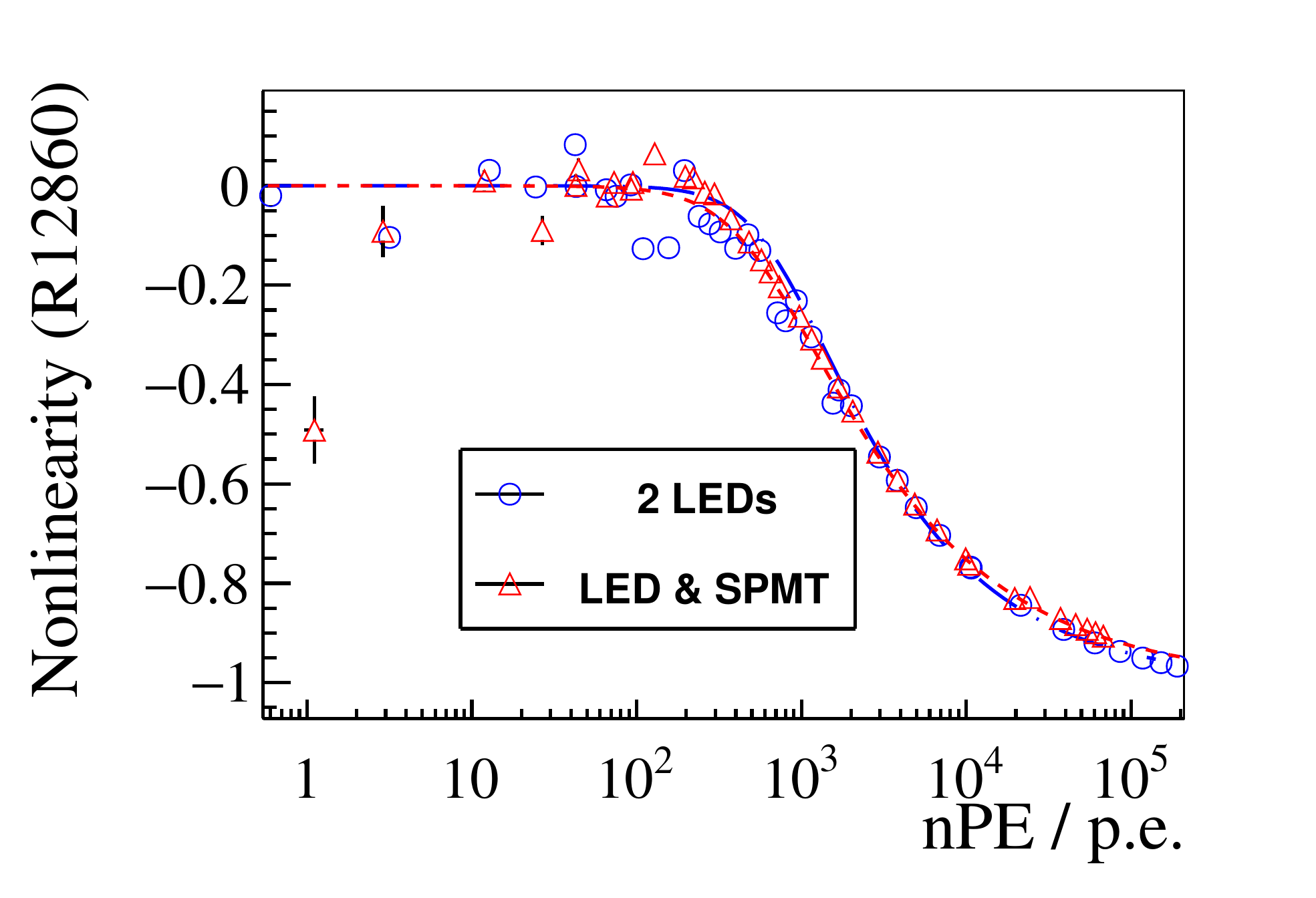}
	\caption{20" dynode-PMT}
	\label{fig:Charge:dynode}
	\end{subfigure}	
	\caption{Charge linearity responses of 20" MCP-PMT and dynode-PMT.}
	\label{fig:Charge}
\end{figure}

The charge linearity of the 20" PMT measured by the two methods is shown in Fig.~\ref{fig:Charge}, where the nonlinearity is defined as the ratio of the difference between the measured and expected charge to the expected charge. The calculation of the relative method and its error propagation are shown in Eq.\ref{eq:relativeMethod} and Eq.\ref{eq:relativeError}, respectively. The $N_{LPMT}$ and $N_{SPMT}$ are the output charge of the LPMT, and the output charge of the SPMT, respectively. The part of $r\times N_{SPMT}$ regarded as the expected charge of LPMT, where the $r$ is the ratio of the charge measured by the LPMT and SPMT by a linear fitting in the range of low light intensity as shown in Fig.\ref{fig:3inch:linearFitting}. The error of $N_{LPMT}$ and $N_{SPMT}$ includes the statistical error and the systemic error, while the error of $r$ is from the fitting, which only includes the statistical error.

\begin{equation}
    n = \frac{N_{LPMT} - r\times N_{SPMT}}{r\times N_{SPMT}}
    \label{eq:relativeMethod}
\end{equation}

\begin{equation}
        \Delta n = \sqrt{\left(\frac{\partial n}{\partial N_{LPMT}}\right)^2\left|\Delta N_{LPMT}\right|^2 + \left(\frac{\partial n}{\partial N_{SPMT}}\right)^2\left|\Delta N_{SPMT}\right|^2 + \left(\frac{\partial n}{\partial r}\right)^2\left|\Delta r\right|^2}
        \label{eq:relativeError}
\end{equation}

The results of double-LED and relative methods are consistent when the expected light intensity is higher than 10\,p.e.~viewed by the 20" PMT. The large uncertainty of the relative method is from the measurement of the 3" PMT, where the signal intensity is too small and comparable to its noise level. The affected range of light intensity is lower than 10\,p.e.~viewed by the 20" PMT considering the factor between the 20" and 3" PMTs.
The nonlinearity curve of the two types of 20" PMTs shows a similar trend as the 3" PMT in Fig.\ref{fig:linearity} and further fits with Eq.~\ref{eq:nonlinearity}\cite{PMTModel}, respectively. 
The $N_{pe}^{sat}$ is $13,548\pm4$\,p.e. for MCP-PMT and $1,221\pm8$\,p.e. for dynode-PMT from the double-LED method.

\subsection{Typical waveform}
\label{2:wave}

\begin{figure}[!htpb]
	\centering
	\begin{subfigure}[c]{0.32\textwidth}
	\centering
	\includegraphics[width=\textwidth]{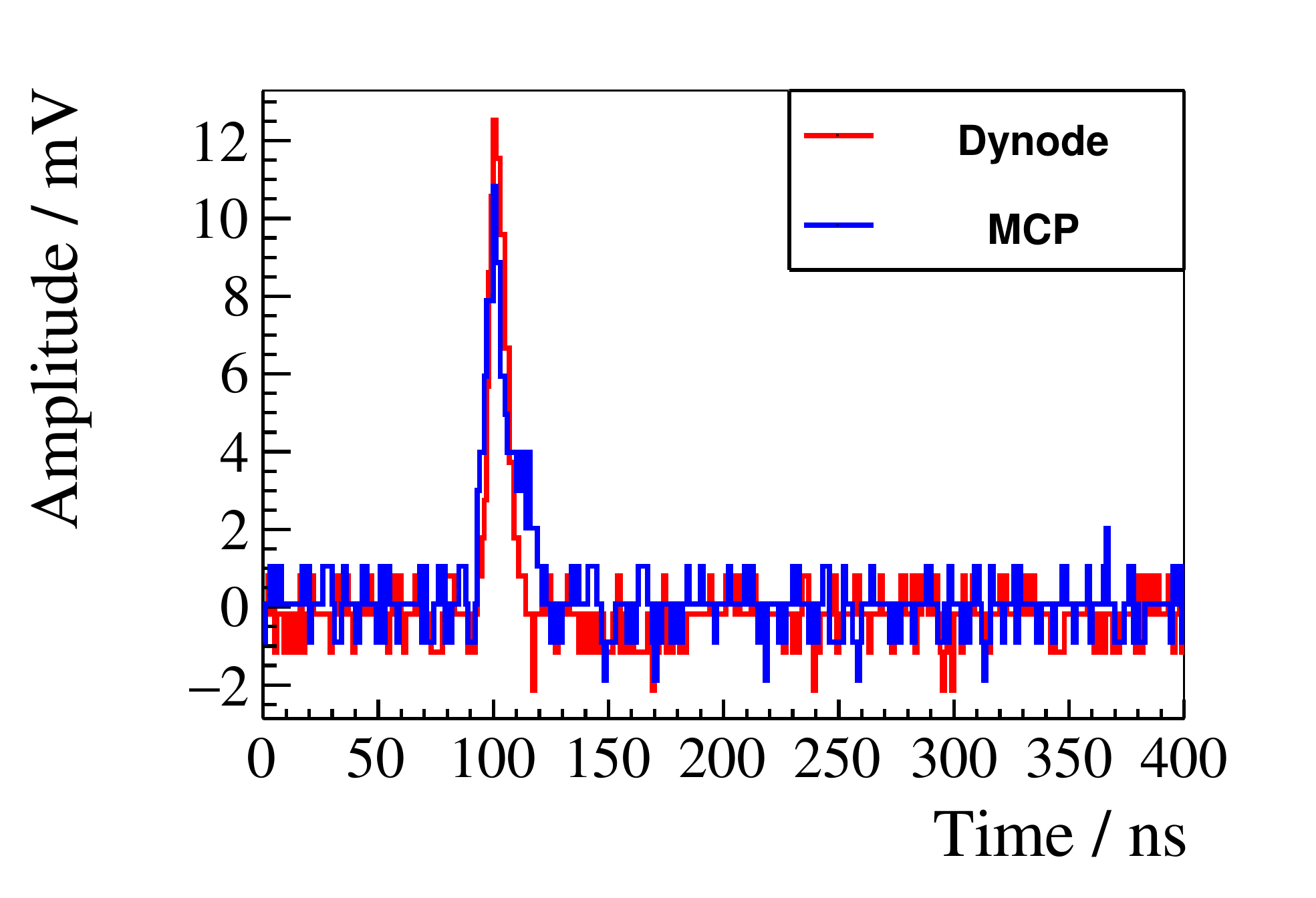}
	\caption{0.5\,p.e.}
	\label{fig:wave:1}
	\end{subfigure}	
	\begin{subfigure}[c]{0.32\textwidth}
	\centering
	\includegraphics[width=\textwidth]{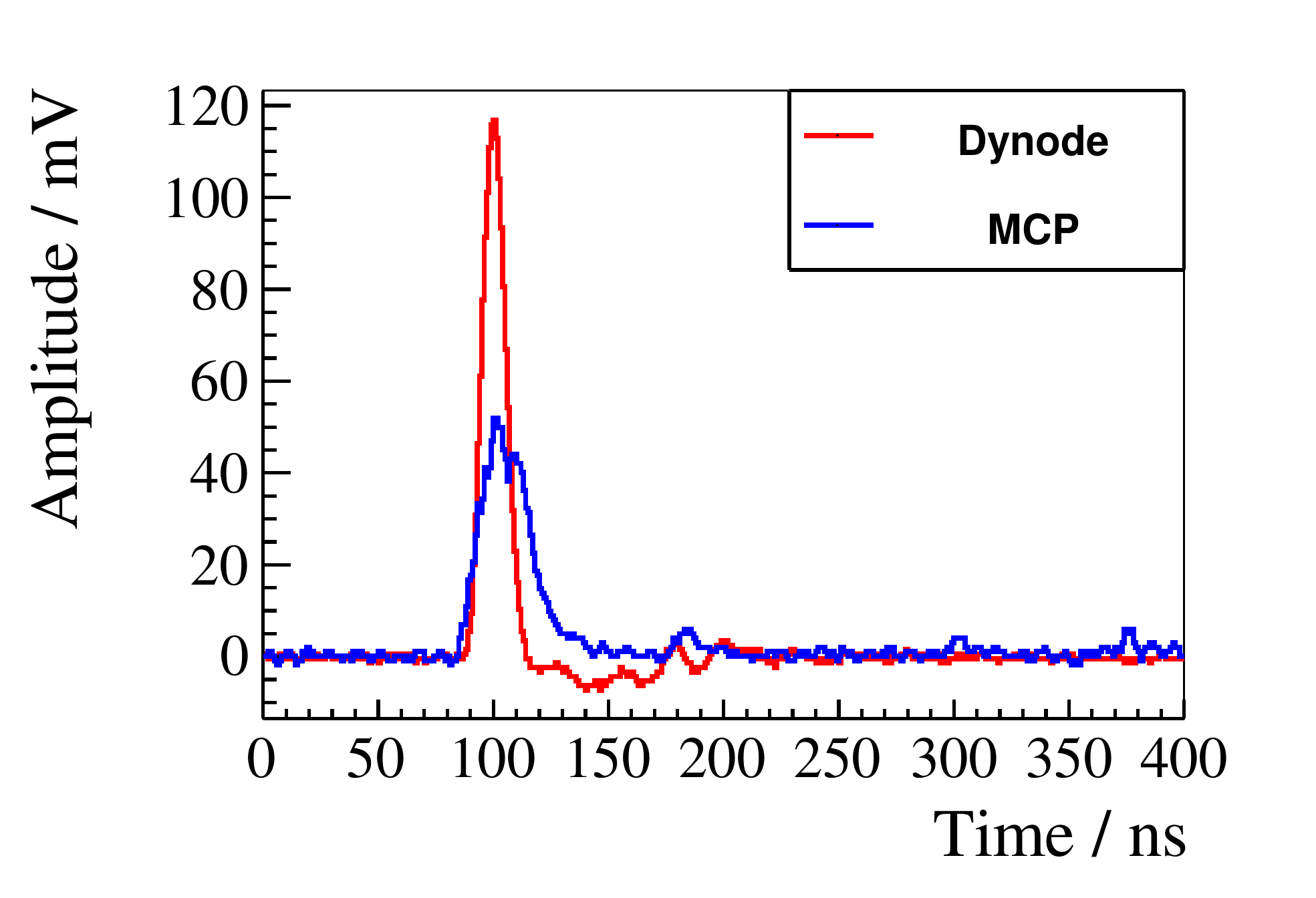}
	\caption{13\,p.e.}
	\label{fig:wave:2}
	\end{subfigure}	
	\begin{subfigure}[c]{0.32\textwidth}
	\centering
	\includegraphics[width=\textwidth]{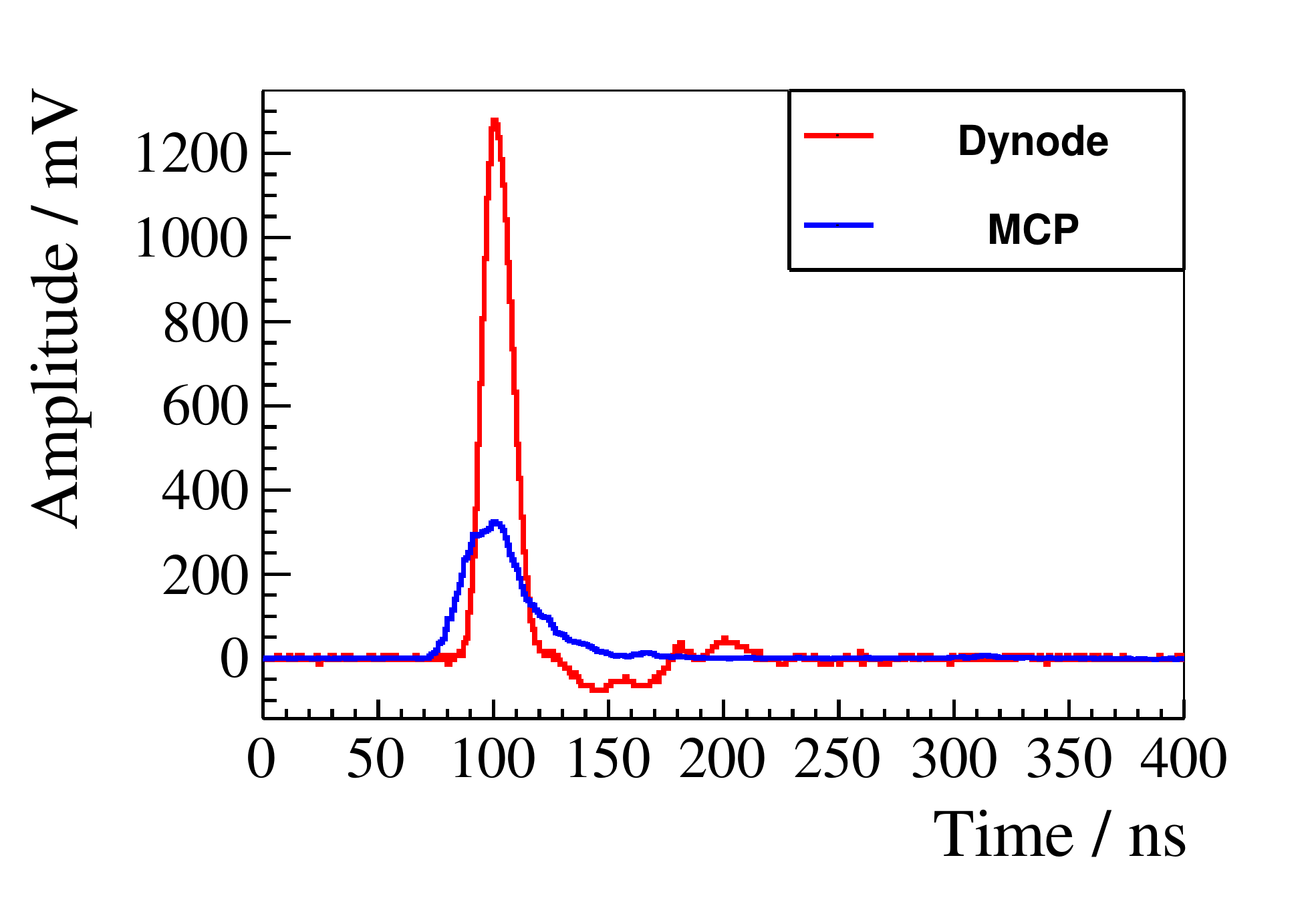}
	\caption{100\,p.e.}
	\label{fig:wave:3}
	\end{subfigure}	
	\begin{subfigure}[c]{0.32\textwidth}
	\centering
	\includegraphics[width=\textwidth]{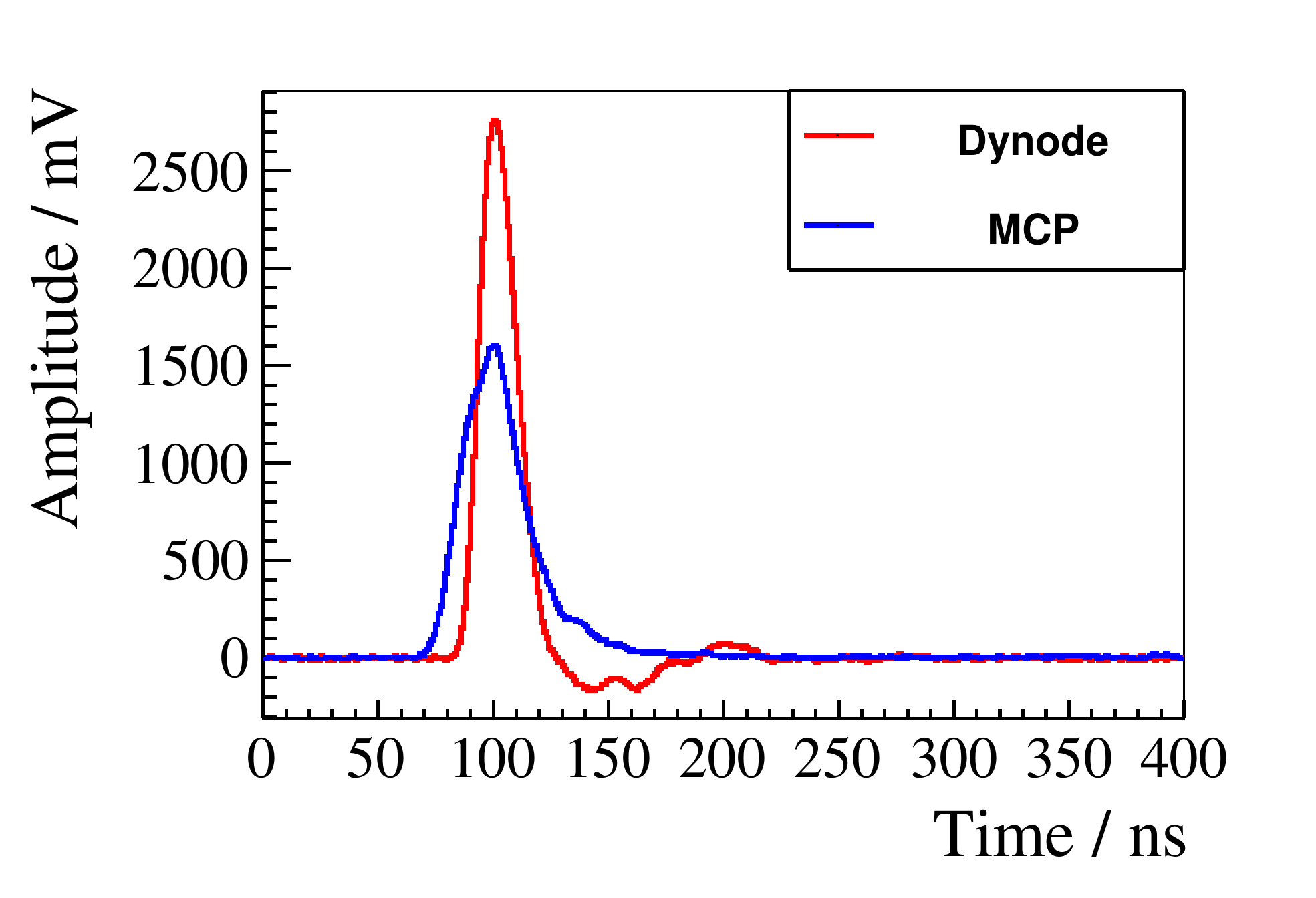}
	\caption{580\,p.e.}
	\label{fig:wave:4}
	\end{subfigure}
	\begin{subfigure}[c]{0.32\textwidth}
	\centering
	\includegraphics[width=\textwidth]{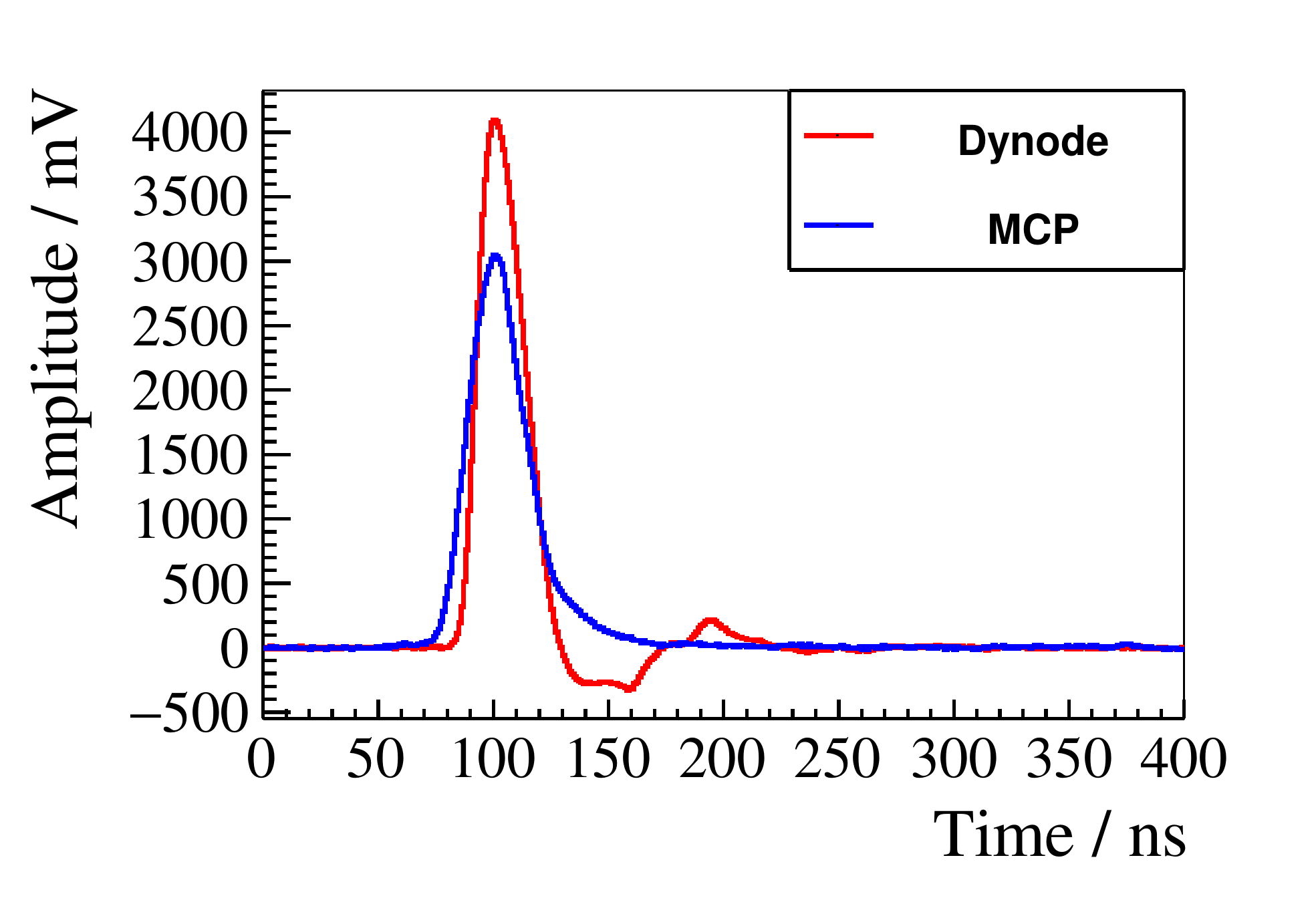}
	\caption{1,000\,p.e.}
	\label{fig:wave:5}
	\end{subfigure}	
	\begin{subfigure}[c]{0.32\textwidth}
	\centering
	\includegraphics[width=\textwidth]{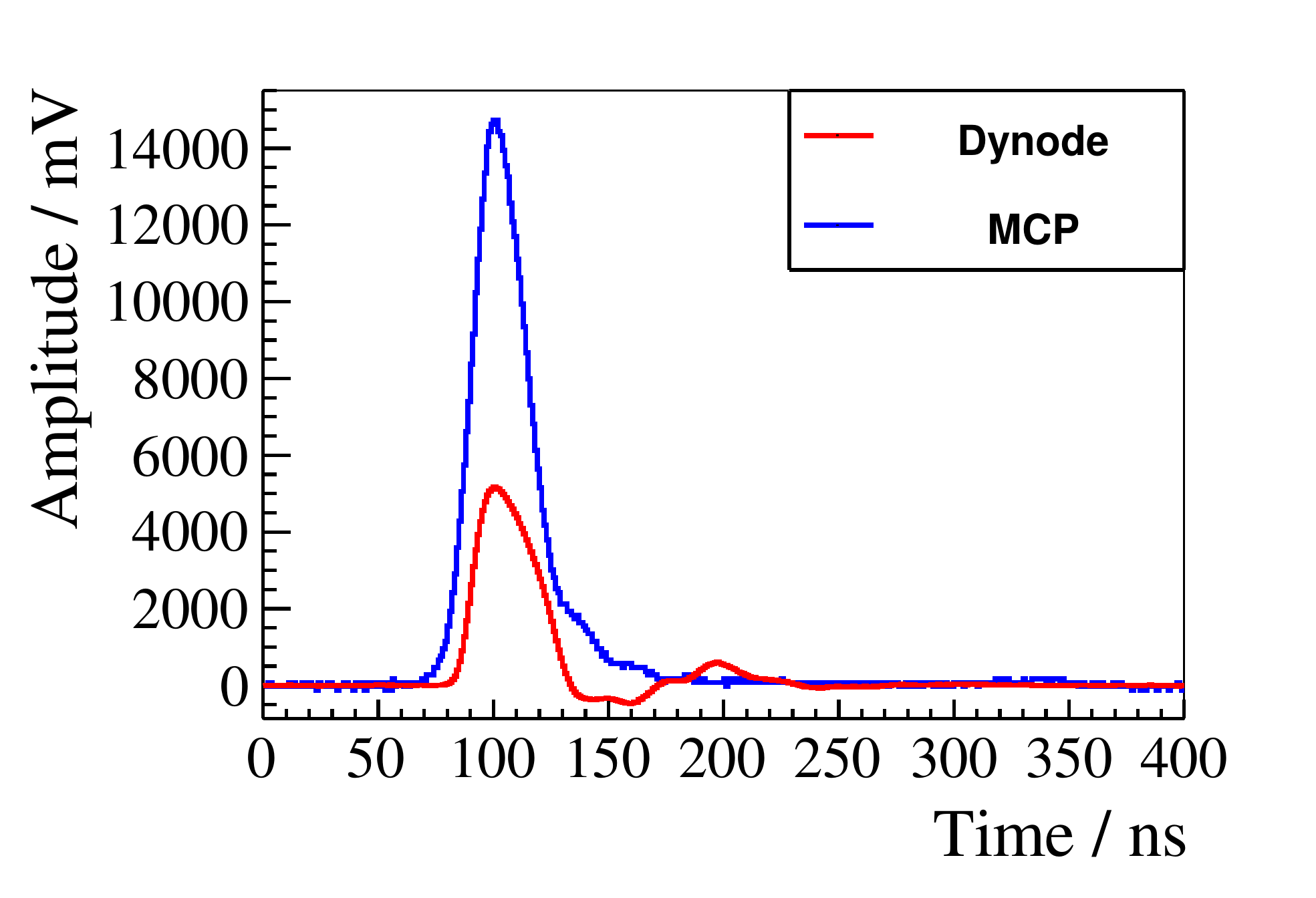}
	\caption{5,000\,p.e.}
	\label{fig:wave:6}
	\end{subfigure}	
	\begin{subfigure}[c]{0.32\textwidth}
	\centering
	\includegraphics[width=\textwidth]{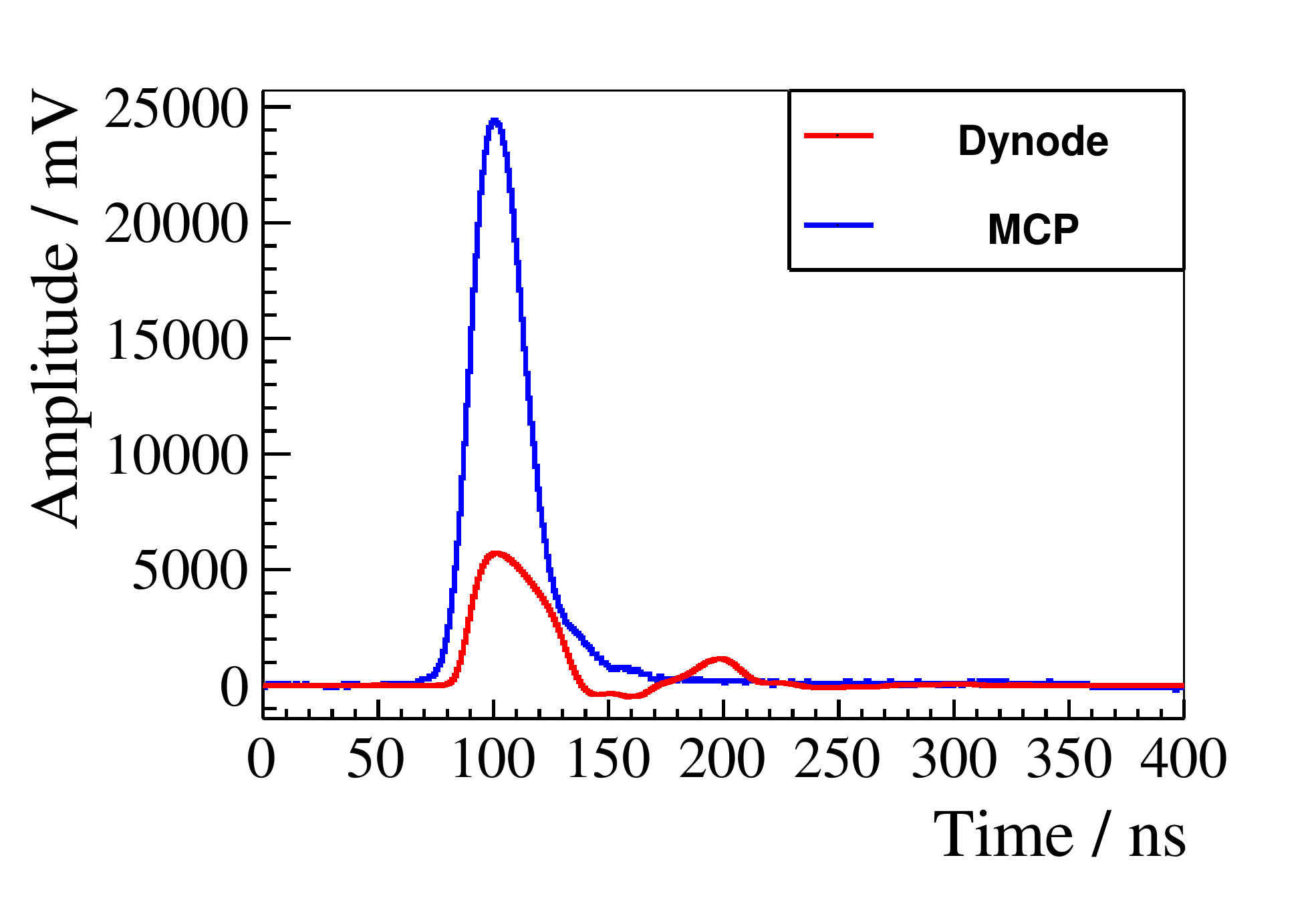}
	\caption{10,000\,p.e.}
	\label{fig:wave:7}
	\end{subfigure}	
	\begin{subfigure}[c]{0.32\textwidth}
	\centering
	\includegraphics[width=\textwidth]{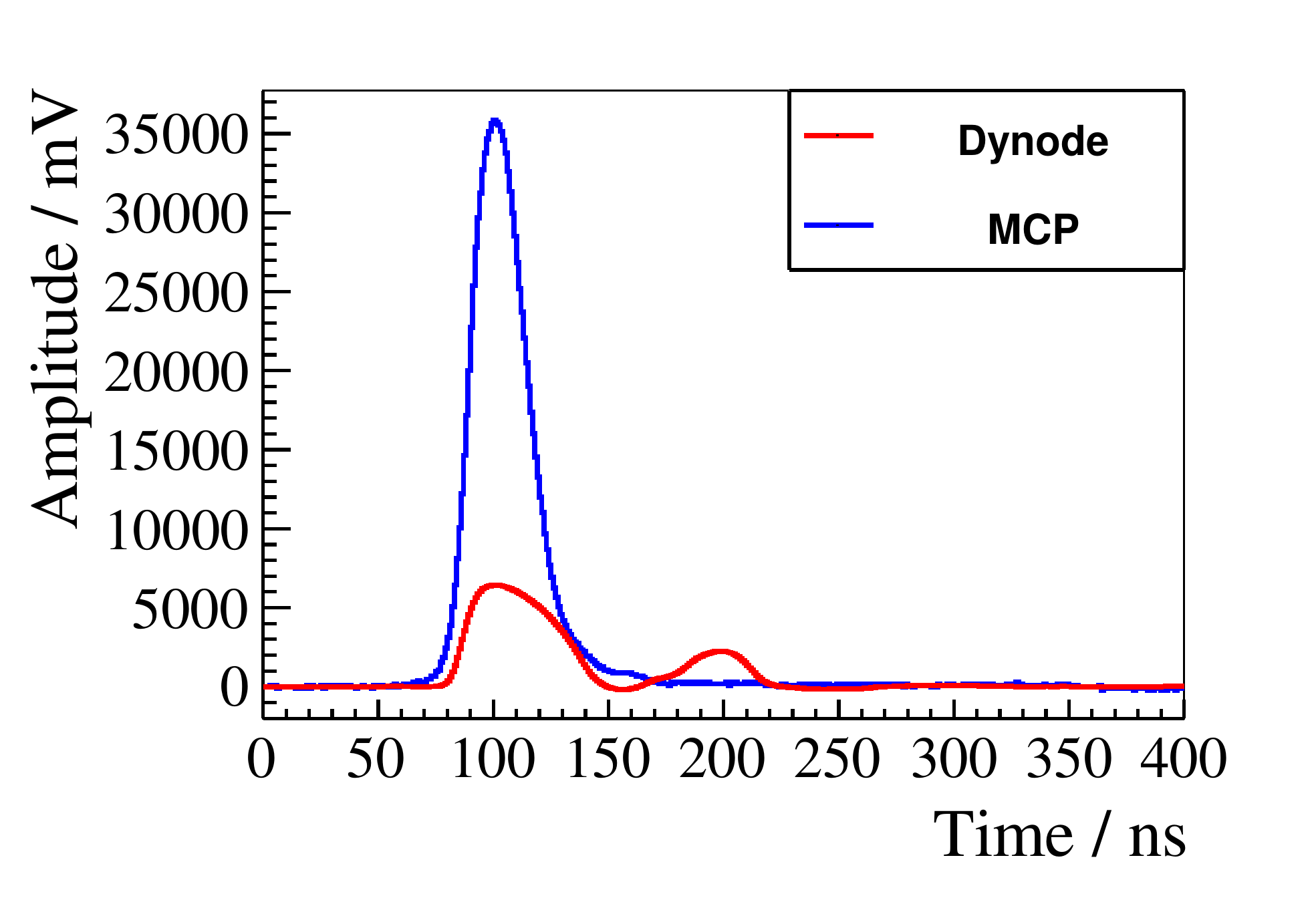}
	\caption{15,000\,p.e.}
	\label{fig:wave:8}
	\end{subfigure}	
	\begin{subfigure}[c]{0.32\textwidth}
	\centering
	\includegraphics[width=\textwidth]{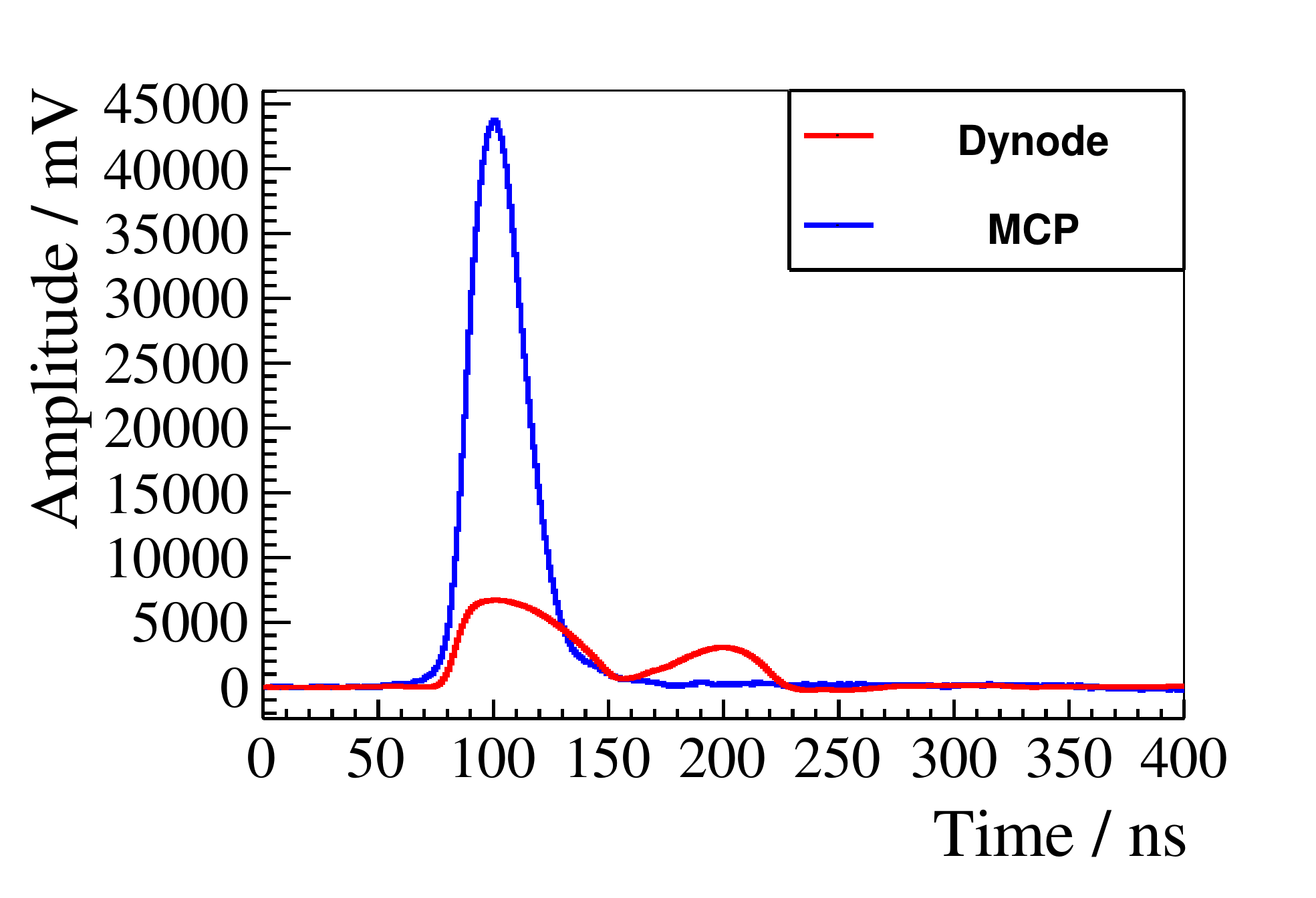}
	\caption{20,000\,p.e.}
	\label{fig:wave:9}
	\end{subfigure}	
	\caption{Typical waveforms with different charge output of 20" PMTs.}
	\label{fig:wave}
\end{figure}

The recorded waveforms of the PMT pulse during the measurements provide more possibilities for a further and better understanding of the PMT linearity features.
Fig.\,\ref{fig:wave} shows the typical waveforms of 20" PMTs with different output charges. We can see that the MCP-PMT shows a more stable pulse shape even with a very large output charge, while the shape of the dynode-PMT pulse is changing obviously when the output charge is increasing. The amplitude of MCP-PMT is still increasing when the expected charge is higher than 10,000\,p.e., while the pulse shape of dynode-PMT is changing obviously in pulse width, overshoot, and late-pulse. The late pulse around 100\,ns after the primary pulse of the 20" dynode-PMT is showing up and enlarging versus the increasing of the output charge when the output charge is larger than 1,000\,p.e., while the MCP-PMT doesn't have this feature. The larger overshoot (few percent in amplitude\cite{fengjiao-PMT-overshoot,fengjiao-PMT-optimization,caimei-container-4}) and reflection are further typical features of the dynode-PMT, which will induce a systematic lower charge according to the defined integration window. The MCP-PMT is not suffered from these problems as seen. 

\subsection{Amplitude}
\label{2:amp}

\begin{figure}[!htpb]
	\centering
	\includegraphics[width=0.8\textwidth]{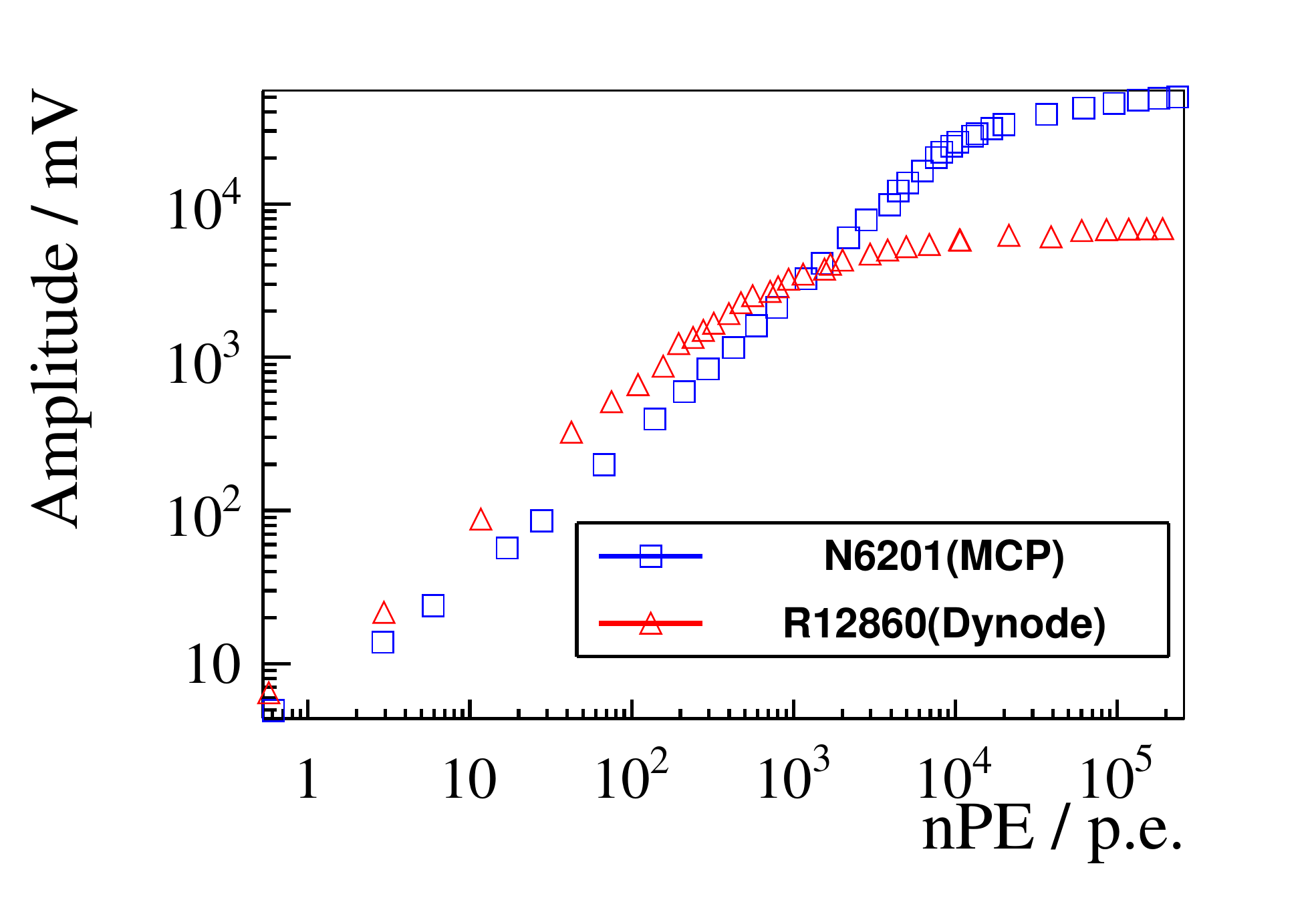}
	\caption{Amplitude versus expected output charge of MCP-PMT and dynode-PMT with the statistical error and the systemic error (too small to see in the plots).}
	\label{fig:Amplitude}
\end{figure}

The pulse amplitude versus the expected output charge of the two types of 20" PMTs is shown in Fig.~\ref{fig:Amplitude}. It is clear that the MCP-PMT shows a wider linear-like response range between the amplitude and charge than the dynode-PMT, where the linear-like range between the charge and amplitude is from 0 to about 8,000\,p.e.~of the MCP-PMT, while it is from 0 to about 300\,p.e.~of the dynode-PMT.

When the expected charge is higher than 100~p.e., the amplitude of MCP-PMT is still linear to the charge, and its amplitude is saturated to around 40-50\,V after 5,000~p.e. The amplitude of dynode-PMT is increasing slower than the MCP-PMT in the range from 100 to 2,000~p.e., then its amplitude is almost saturated to around 7\,V. The larger amplitude of MCP-PMT could be sourced from the internal capacitor of the MCP itself and higher electric field strength because of its thickness dimension on a sub-millimeter scale than the dynode.

\begin{figure}
    \centering
    \begin{subfigure}[c]{0.45\textwidth}
	\centering
    \includegraphics[width=\textwidth]{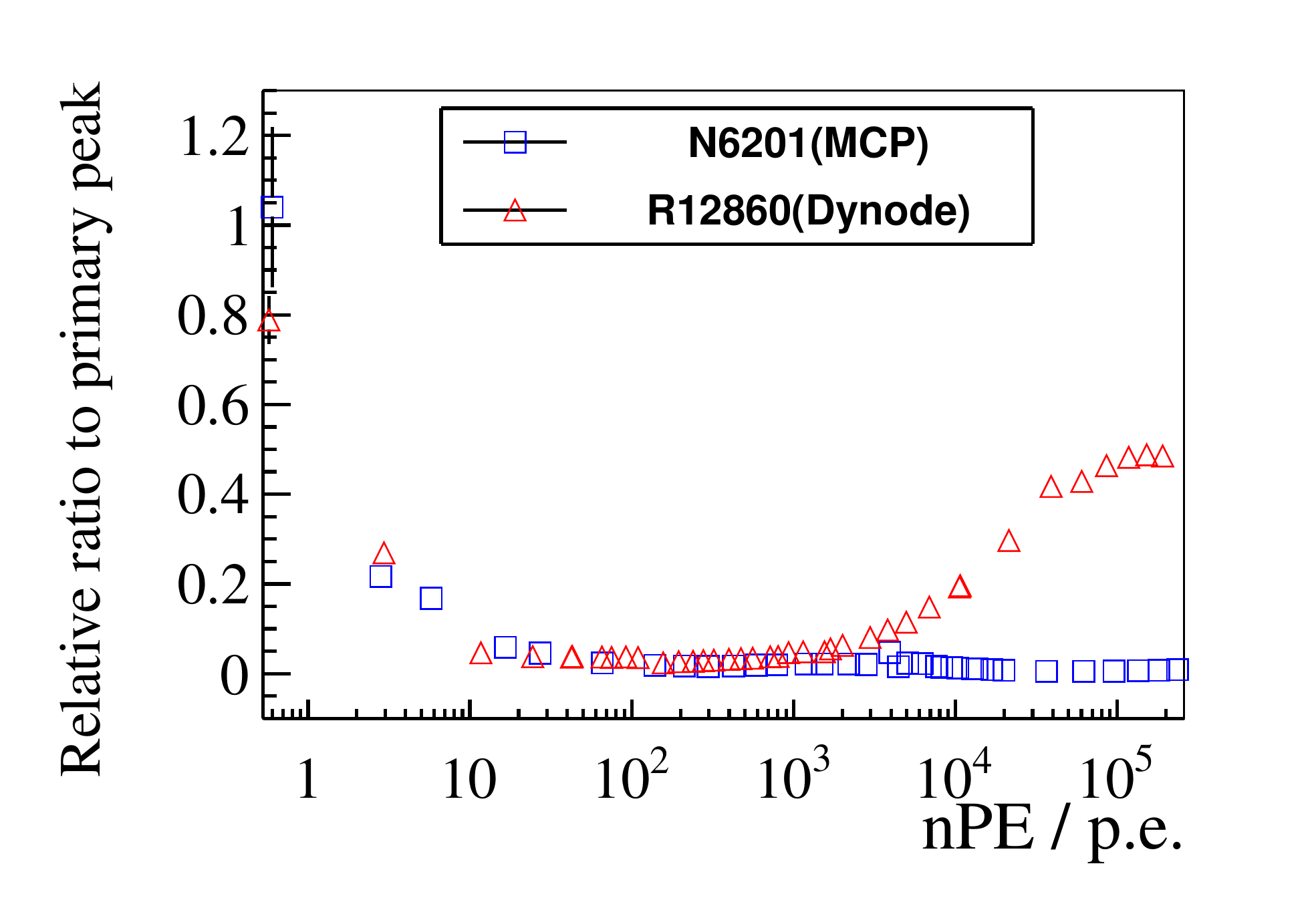}
    \caption{The amplitude ratio of the late-pulse.}
    \label{fig:laterpulse}
    \end{subfigure}
    \begin{subfigure}{0.45\textwidth}
    \centering
    \includegraphics[width=\textwidth]{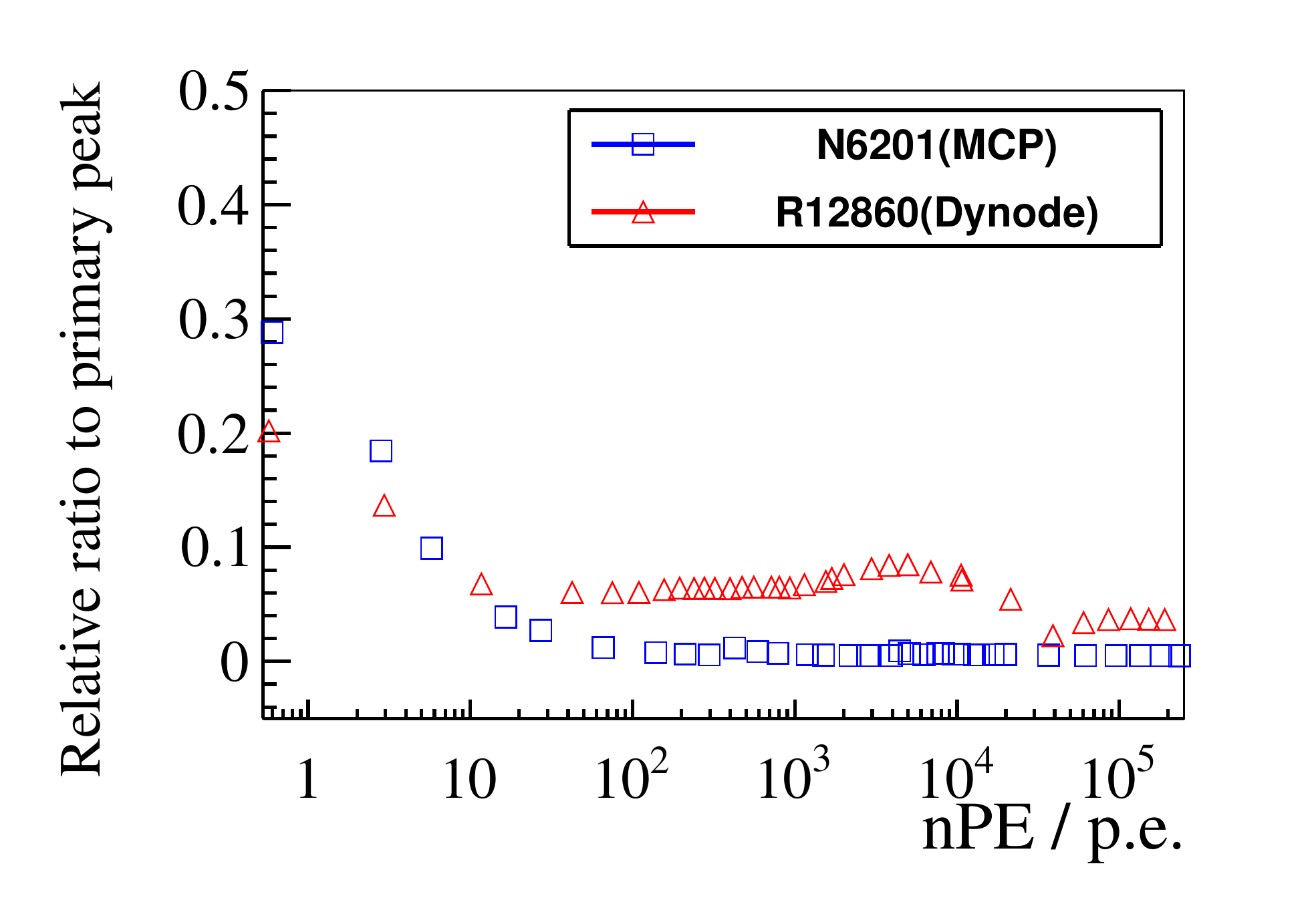}
    \caption{The amplitude ratio of the overshoot.}
    \label{fig:overshoot}
    \end{subfigure}
    \caption{The amplitude ratio of the late pulse and overshoot to the primary pulse, where the late pulse is located at around 100\,ns after its primary pulse, and the overshoot is the minimum amplitude after its primary pulse.}
    \label{fig:latepulse:overshoot}
\end{figure}

The amplitude of the late pulse, around 100\,ns after the primary pulse, is also derived from the data, and its amplitude ratio to the primary pulse is shown in Fig.\,\ref{fig:laterpulse}. The late pulse of the dynode-PMT is increasing to 50\% of the primary pulse when the expected charge is higher than around 2,000\,p.e. The ratio is dominated by the noise and the reflection when the charge of the primary pulse is less than around 10\,p.e. The late pulse of the dynode-PMT will affect the charge and fall-time calculation of the primary pulse. The MCP-PMT shows a very different feature to the dynode-PMT on the late pulse, which is mainly dominated by the reflection of the primary pulse at a few percent levels. The amplitude ratio of the overshoot to its primary pulse is shown in Fig.\,\ref{fig:overshoot}. The dynode-PMT suffers a larger overshoot than the MCP-PMT, and both of the curves are dominated by the noise when the expected charge is smaller than 10\,p.e.

\subsection{Timing-related parameters}
\label{2:Time}

\begin{figure}[!htpb]
	\centering
	\begin{subfigure}[c]{0.45\textwidth}
	\centering
	\includegraphics[width=\textwidth]{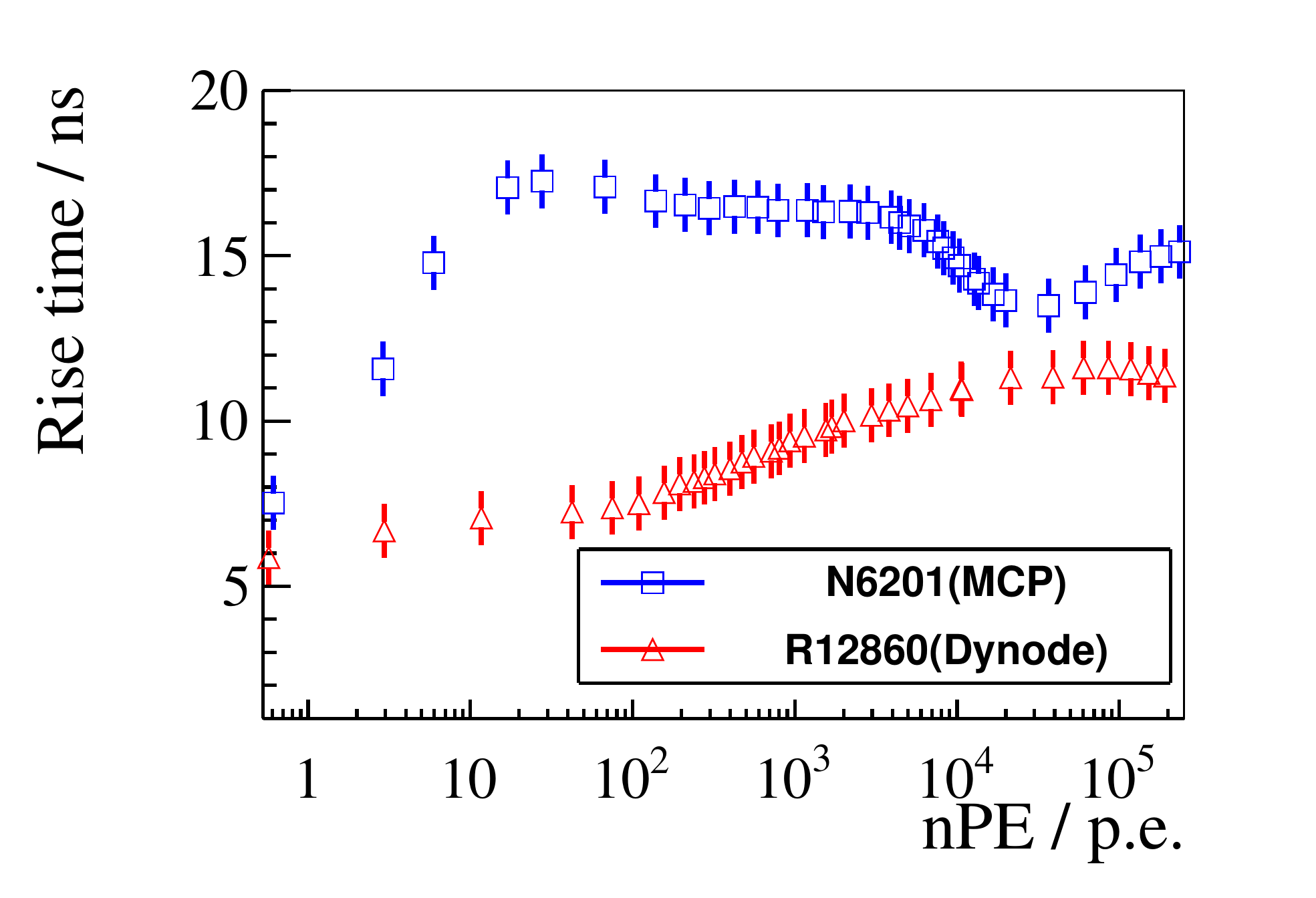}
	\caption{Rise-time}
	\label{fig:timing:rise}
	\end{subfigure}	
	\begin{subfigure}[c]{0.45\textwidth}
	\centering
	\includegraphics[width=\textwidth]{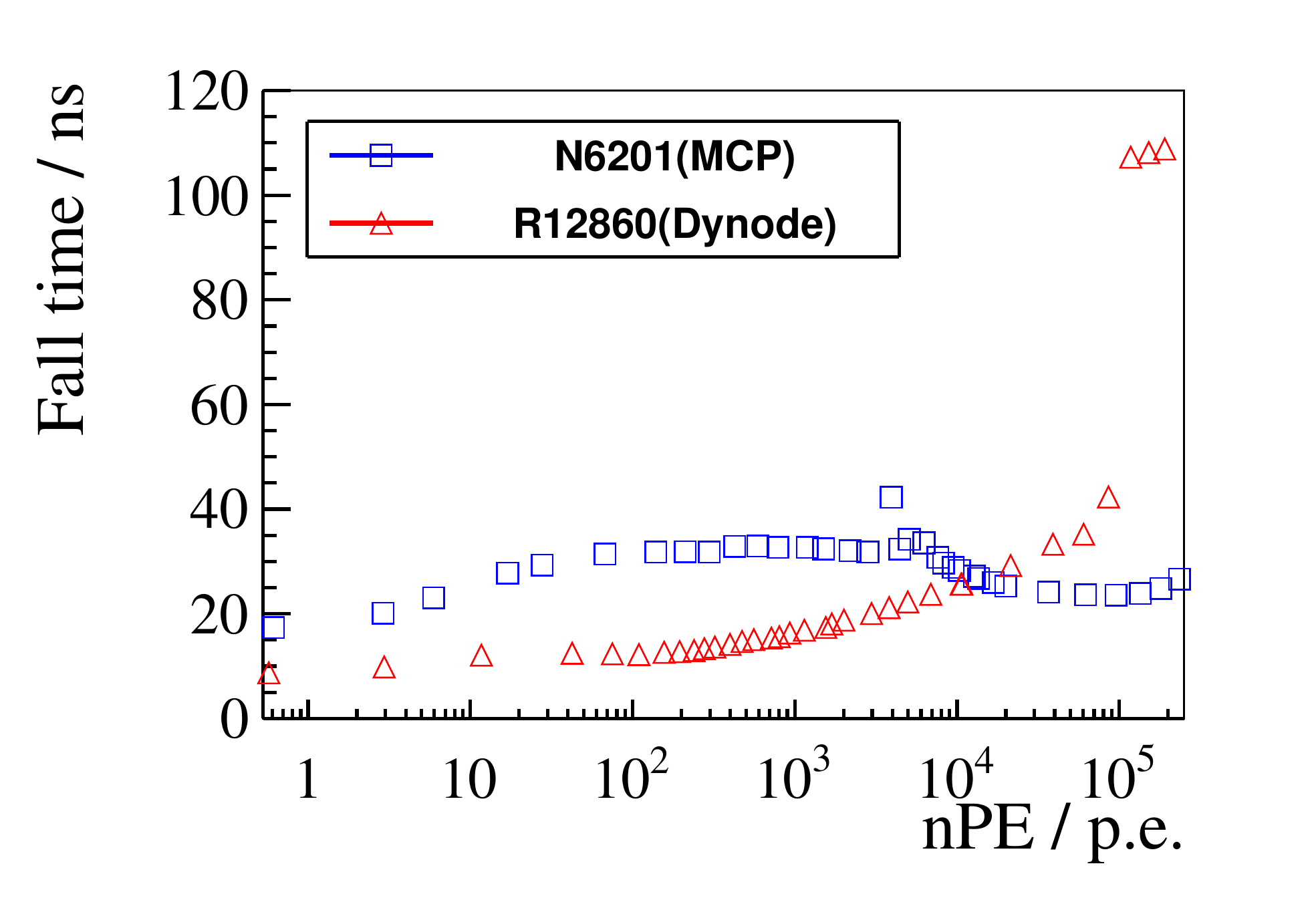}
	\caption{Fall-time}
	\label{fig:timing:fall}
	\end{subfigure}	
	\begin{subfigure}[c]{0.45\textwidth}
	\centering
	\includegraphics[width=\textwidth]{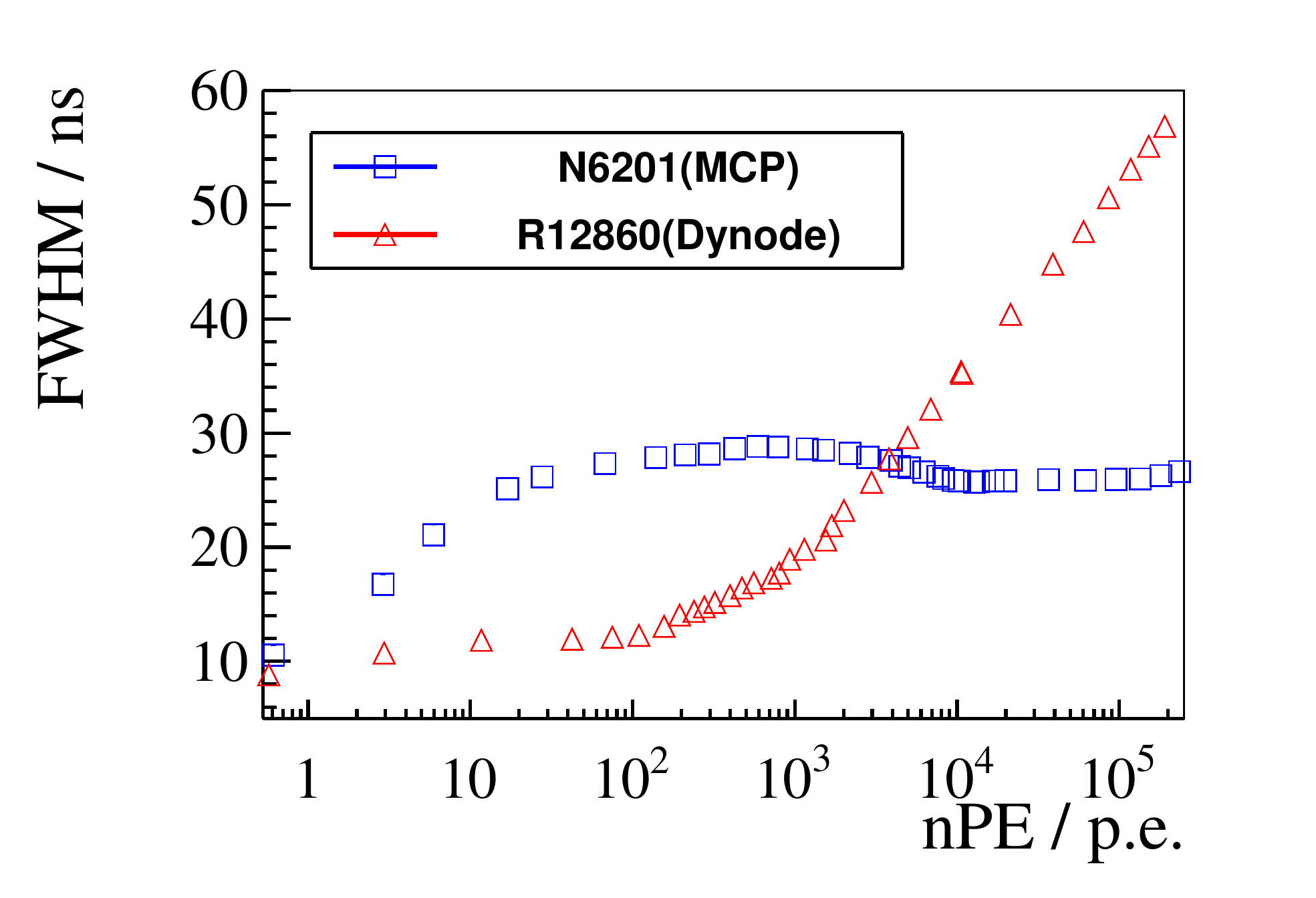}
	\caption{FWHM}
	\label{fig:timing:fwhm}
	\end{subfigure}	
  	\begin{subfigure}[c]{0.45\textwidth}
 	\centering
 	\includegraphics[width=\textwidth]{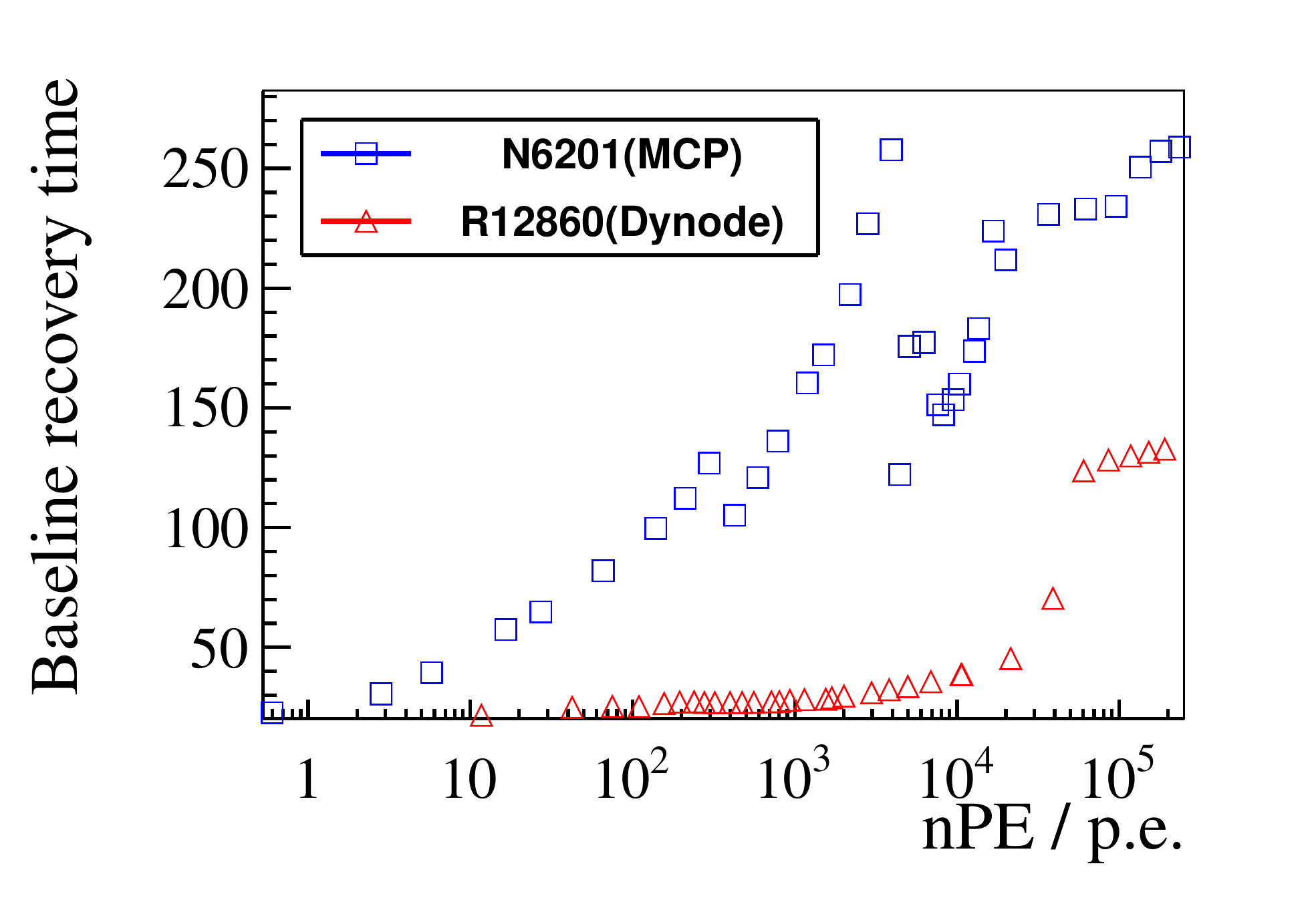}
 	\caption{Baseline-recovery time}
 	\label{fig:timing:tobasetime}
 	\end{subfigure}	
	\caption{Timing parameters versus expected output charge.}
	\label{fig:timing}
\end{figure}

The parameters related to PMT timing are shown in Fig.~\ref{fig:timing}, where the two types of PMTs show very different features as expected from the typical waveforms too. It is worth mentioning that the large deviation of dynode PMT's last several points in the figures of fall-time and baseline-recovery time are due to the unignorable later pulse.

The rise-time of MCP-PMT increases quickly from 7\,ns at 1\,p.e. to 17\,ns at around 20\,p.e., then slowly goes down anti-correlated to the expected charge increasing, and a valley around 20,000\,p.e. But the rise-time of dynode-PMT is increasing slowly only from 5\,ns to around 10\,ns. 
The fall-time of MCP-PMT increases quickly from 18\,ns at around 1\,p.e. to 30\,ns at around 100\,p.e., keeping in constant till to thousands of p.e.s, and a valley structure around 30,000\,p.e. A point of the fall-time curve of the MCP-PMT is out of the trend and pretty larger than others at about 2800\,p.e., which is further checked and normal in data, and is a possible source from the light source. The fall-time of dynode-PMT keeps constant till to around 100\,p.e., and increases quickly after 1,000\,p.e.\,to around 40\,ns. The jump around 100,000\,p.e.~of dynode-PMT is from the late pulse, which is merging into the primary pulse. 
The FWHM of MCP-PMT increases quickly from 10\,ns around 1\,p.e. to 25\,ns at around 20\,p.e., keeping almost in constant to around few 1,000s\,p.e., then going down following the expected charge increasing and keep in constant again after around 10,000\,p.e. The FWHM of dynode-PMT is keeping in constant around 11\,ns to 100\,p.e., then increasing quickly from 20\,ns to around 50\,ns after 10,000\,p.e. 
The baseline-recovery time after the primary pulse is derived from the data too. The baseline of the dynode-PMT after the primary pulse recovers less than 50\,ns when the expected output charge is less than 10,000\,p.e., and it is enlarged to around 150\,ns by the merged late-pulse. The recovery time of the MCP-PMT baseline is increasing following the expected output charge increase and shows fluctuation in a valley around 5,000\,p.e.

In short, the pulse shape of MCP-PMT is getting a little bit wider and has a larger amplitude than the dynode-PMT when the expected output charge is increasing, while that of dynode-PMT is getting much wider with a saturated amplitude when the expected charge is higher than several thousand p.e. 

\section{Summary}
\label{1:sum}
With the detailed measurements and comparison of two linearity calibration methods, the relative method is working well as the traditional double-LED method in pulse mode. 
The relative method makes it easier to calibrate the linearity response of 20" PMT in the experiment. This method can be implemented in many experiments, even in JUNO.
The features including charge, amplitude, late pulse, and time-related features of two kinds of 20" PMTs are also compared and found that they are pretty different. The 20" MCP-PMT has a large range (more than 10,000~p.e.) of linear response and can reach a huge amplitude (50~V) with a narrow pulse shape (FWHM around 26~ns), while the 20" dynode-PMT responses in nonlinearity with saturated amplitude (around 7\,V) after only about 1,000~p.e., and the dynode PMT shows a wider pulse shape (FWHM more than 30~ns) as light intensity is large.


\section*{Acknowledgments}
\label{sec:1:acknow}

This work was supported by the National Natural Science Foundation of China (NSFC) No. 11875282, 11475205, and 12042507, the Strategic Priority Research Program of the Chinese Academy of Sciences (Grant No. XDA100102).

\bibliographystyle{unsrtnat}
\bibliography{allcites}   

\end{document}